\documentclass[journal]{IEEEtran}

\usepackage{xcolor}
\usepackage{cancel}
\usepackage{dblfloatfix}
\usepackage{ragged2e}
\usepackage{graphicx}
\usepackage{booktabs}
\usepackage{makecell}
\usepackage{multirow}
\usepackage{pifont}
\usepackage{multicol}
\usepackage{wrapfig}
\usepackage{subcaption}
\usepackage{url}
\usepackage{pgf-pie}
\usepackage{pgfplots}
\usepackage{xspace}
\usepackage{lstautogobble}
\usepackage[labelfont=bf]{caption}
\usepackage{todonotes}
\usepackage{titlesec}
\usepackage{tabularx}
\titlespacing*{\paragraph}{0pt}{0pt}{1em}
\titleformat{\paragraph}[runin]{\normalfont\normalsize\bfseries}{}{0pt}{}[:]
\usepackage{enumitem}
\setlist[enumerate]{left=0pt}

\newcommand{\ignore}[1]{}

\def \tool{\textsc{CaSey}\xspace}

\begin{document}

\title{Streamlining Security Vulnerability Triage with Large Language Models
}

\author{
    \IEEEauthorblockN{Mohammad Jalili Torkamani
    \IEEEauthorrefmark{1}, Joey Ng
    \IEEEauthorrefmark{1}, Nikita Mehrotra
    \IEEEauthorrefmark{2}, Mahinthan Chandramohan
    \IEEEauthorrefmark{3}, Padmanabhan Krishnan
    \IEEEauthorrefmark{3}, Rahul Purandare
    \IEEEauthorrefmark{1}}\\
    \IEEEauthorblockA{
        \IEEEauthorrefmark{1}School of Computing, University of Nebraska--Lincoln, Lincoln, USA \\
        Email: mJaliliTorkamani2@huskers.unl.edu, jng4@huskers.unl.edu, rahul@unl.edu}\\
    \IEEEauthorblockA{
        \IEEEauthorrefmark{2}Microsoft, New Delhi, India \\
        Email: nmehrotra@microsoft.com}\\
    \IEEEauthorblockA{
        \IEEEauthorrefmark{3}Oracle Labs, Brisbane, Australia \\
        Email: mahin.chandramohan@oracle.com, paddy.krishnan@oracle.com}
}

\maketitle

\begin{abstract}
Bug triaging for security vulnerabilities is a critical part of software maintenance, ensuring that the most pressing vulnerabilities are addressed promptly to safeguard system integrity and user data. However, the process is resource-intensive and comes with challenges, including classifying software vulnerabilities, assessing their severity, and managing a high volume of bug reports. In this paper, we present \tool, a novel approach that leverages Large Language Models (in our case, the GPT model) that automates the identification of Common Weakness Enumerations (CWEs) of security bugs and assesses their severity. \tool employs prompt engineering techniques and incorporates contextual information at varying levels of granularity to assist in the bug triaging process. We evaluated \tool using an augmented version of the National Vulnerability Database (NVD), employing quantitative and qualitative metrics to measure its performance across CWE identification, severity assessment, and their combined analysis. \tool achieved a CWE identification accuracy of 68\%, a severity identification accuracy of 73.6\%, and a combined accuracy of 51.2\% for identifying both. These results demonstrate the potential of LLMs in identifying CWEs and severity levels, streamlining software vulnerability management, and improving the efficiency of security vulnerability triaging workflows.
\end{abstract}


\section{Introduction and Motivation}
\label{sec:introduction}

Software bugs are inherent challenges in the development and maintenance of software systems \cite{rodriguez2020bugs}. These flaws range from minor functional issues to severe security vulnerabilities with far-reaching consequences \cite{li2021comprehensive}. One notable example is the 2017 Equifax data breach, where a vulnerability in a widely used software framework exposed the personal information of around 145 million individuals \cite{equifax2017}. This incident not only inflicted significant reputational damage and financial losses but also highlighted the critical need for rigorous and proactive vulnerability management. Therefore, the accurate identification of bug types, identifying their severities, and timely resolution are vital to maintaining system integrity, minimizing risks, and fostering user trust in an increasingly interconnected digital world \cite{ahsan2009automatic}.

Bugs can originate from diverse factors, including source code, component interactions, system integration, and environmental setups \cite{rodriguez2020bugs}. Security-related vulnerabilities are particularly concerning. These flaws can enable attackers to exploit systems, resulting in data breaches, unauthorized access, denial of service, or full system compromise \cite{aslan2023comprehensive}. The ever-evolving digital landscape amplifies the threat as malicious actors continuously refine their methods \cite{bendovschi2015cyber}. 

Bug triaging \cite{murphy2004automatic,10.1145/1985793.1985934} is the systematic process of evaluating, prioritizing, and assigning bugs to appropriate developers for resolution. This practice ensures that the most critical issues are addressed promptly, while less urgent ones are handled later. Ideally, bugs are triaged based on their urgency \cite{almhana2021considering}. However, manual triaging, while reliable, is time-consuming and often relies on human judgment, which can be prone to biases, inconsistencies, and inefficiencies. Moreover, as the size and complexity of software systems grow, the number of bugs can become overwhelming, making it increasingly difficult to manage effectively without automated assistance \cite{10.1145/3238147.3238213}. For security-related bugs, timely identification and prioritization are especially crucial, as they directly impact the integrity and safety of systems and users \cite{walkowski2021vulnerability}.

To overcome these limitations, recent advancements in computer science and software engineering, particularly in machine learning, have shown promise in efficiently capturing these bugs, offering more accurate and efficient methods of bug triaging and severity identification \cite{5298419, alazzam2020automatic, liu2022automatic, nagwani2023artificial}. While semi-automated tools assist in managing bug triaging, they come with inherent limitations. For example, BugListener \cite{9793897} synthesizes reports from live chat data, which presents challenges related to data privacy and adaptability. Such reliance on specific data sources limits the scalability and flexibility of these tools, highlighting the need for more adaptable, automated solutions capable of functioning without proprietary data constraints. Therefore, the mentioned issues, combined with the limitations of existing research, motivate our study to explore Large Language Models (LLMs) \cite{10.1145/3641289, kasneci2023chatgpt} as a potential solution to these challenges. They are capable of processing and understanding vast amounts of structured and unstructured data. Unlike rule-based systems, LLMs can interpret complex and nuanced bug descriptions, facilitating the assignment of bugs to the right developers according to their expertise. Additionally, LLMs can adapt to emerging vulnerabilities and threats \cite{lu2024grace, guo2024outside}, continuously learning from new data to stay up-to-date with evolving security risks. This dynamic learning capability, combined with the LLMs' scalability and consistency, positions them as a powerful tool for software vulnerability identification, automating bug triaging, and improving the overall security posture of software systems \cite{10456393}.

In this paper, we propose \tool, a novel approach leveraging LLMs to identify the origins (Common Weakness Enumeration) and severity of security bugs. Our method employs prompt engineering to analyze bug descriptions and optionally their associated buggy code at various granularities, including files, methods, or hunks\footnote{A hunk refers to a block of changes in a code file to represent a group of added, removed, or modified lines in a code difference.} to understand the impact of each combination. By identifying CWEs (e.g., CWE-416) and severity labels (e.g., HIGH) based on the Common Vulnerability Scoring System (CVSS), \tool enables enterprises to prioritize and assign critical bugs efficiently, ensuring timely resolution of security vulnerabilities.

We fine-tuned the GPT-3.5 model (hereafter referred to as GPT-3) on an enhanced NVD dataset encompassing various levels of code granularity. As shown in Section \ref{sec:results}, this fine-tuning process significantly enhanced \tool's accuracy in identifying CWEs and severity levels. We compare the experimental results to those obtained when using the standard GPT-3 model. GPT-3 was selected due to its demonstrated versatility in understanding and generating high-quality responses across technical and code-related contexts, combined with its cost-effectiveness. Furthermore, the model’s training data, which extends up to September 2021, provides an opportunity to evaluate its ability to generalize to unseen data. This generalization capability is critical for adapting to emerging vulnerabilities and novel security issues that were not part of the model’s training set. This cut-off date was applied when collecting the fine-tuning dataset, ensuring that all records selected for fine-tuning are unseen by the model and distinct from the evaluation dataset (whose records are totally unseen and after the cut-off date). This approach enhances the generative model's capability while ensuring no overlap between the evaluation data and the fine-tuning dataset. Further details will be provided in Section \ref{sec:experiments} of this paper.

Our approach is designed for internal deployment within enterprises, ensuring that sensitive data remains secure while leveraging the tool’s capabilities to identify security vulnerabilities. By hosting the tool that adopts the fine-tuned model on enterprises' infrastructure, enterprises can guarantee that proprietary codebases, confidential vulnerability reports, and other sensitive information do not leave their controlled environment. This setup allows enterprises to effectively utilize the tool for identifying CWEs and assessing severity, all while maintaining full control over data and ensuring compliance with internal policies or regulatory standards. Additionally, internal deployment enables customization to incorporate organization-specific knowledge, further enhancing the tool’s accuracy and relevance for identifying vulnerabilities unique to their software ecosystem. As such, our approach is an effective solution for enterprises aiming to modernize their security practices while protecting intellectual property and sensitive information.

Our key contributions are as follows: 

\begin{enumerate}
\item \textbf{Automating CWE and severity identification:} We introduce \tool to automate CWE and severity identification, streamlining the bug triaging process.
\item \textbf{Dataset Collection:} We curated a dataset comprising 6,031 real-world projects developed in various programming languages, enhancing the National Vulnerability Database (NVD) with buggy code collected from GitHub at varying granularity levels. We Used half of this dataset to fine-tune the LLM and evaluated it on the remaining half, resulting in a specialized model used by our tool.
\end{enumerate}

This paper addresses the following key research questions: 
 \begin{enumerate} 
 \item \textbf{Can \tool accurately identify CWEs?} We will outline the steps taken by our approach to infer CWEs, providing a detailed explanation of the methodology used to design and execute the experiments, along with the resulting outcomes. 
 \item \textbf{Can \tool accurately identify severities based on their associated CVSS version?} We will describe how \tool identifies the severity (both label and score) of a security vulnerability according to its given CVSS version. Similar to the previous RQ, we will also present the experimental setup, results, and evaluation.
 \item \textbf{Can \tool accurately identify CWEs and severities collectively?} By integrating CWE and severity identifications in alignment with their corresponding CVSS version, we will analyze the combined performance of \tool. This analysis will provide deeper insights into how well \tool can be used in security bug triaging on both their CWE category and severity levels. \end{enumerate}

The paper has been organized as follows:

In Section \ref{sec:background}, we present the foundational concepts that underpin the paper. Section \ref{sec:approach} outlines the methodology adopted by our approach, while Section \ref{sec:implementation} provides a detailed description of \tool's construction and its implementation. Section \ref{sec:experiments} covers the dataset collection process, the design of the experiments, and the analysis performed to evaluate the effectiveness of our approach. In Section \ref{sec:results}, we interpret the results of the experiments, identifying patterns and the accuracy of the tool. We also conduct some manual analysis to gain deeper insights into the performance of our approach. In Section \ref{sec:threats_to_validity}, we discuss potential threats to the validity of our findings. Section \ref{sec:related_works} reviews related work in bug triaging, localization, and classification. Finally, Section \ref{sec:conclusion} summarizes the key findings and their implications and suggests directions for future work. At the end of this paper, we provide an access link to the research artifacts, including datasets, experiment results, and the source code for \tool.
\section{Background}
\label{sec:background}
\subsection{Common Weakness Enumeration (CWE):} 
The Common Weakness Enumeration (CWE) \cite{wu2015they} is a comprehensive catalog of known software weaknesses and vulnerabilities, serving as a standardized framework for identifying and categorizing software flaws. Developed and maintained by the MITRE Corporation, CWE is an essential resource for developers, software maintainers, and organizations aiming to enhance their security practices. It aids in identifying and mitigating vulnerabilities, preventing potential exploitation by attackers. Each CWE entry provides detailed insights, including a description of the weakness, illustrative examples when available, and the potential consequences of the vulnerability. Examples of well-documented weaknesses include cross-site scripting (CWE-79), SQL injection (CWE-89), and buffer overflows (CWE-120). The catalog is hierarchically organized, with broad categories of weaknesses subdivided into more specific types. This structure facilitates a systematic understanding of software vulnerabilities and supports targeted remediation efforts. CWE entries often integrate with other security standards, such as the Common Vulnerabilities and Exposures (CVE) system, to deliver a holistic perspective on software security. Organizations leveraging CWE can proactively address vulnerabilities during development, leading to more secure, stable, and reliable systems. By standardizing the classification and communication of software weaknesses, CWE empowers the software community to collaboratively improve security across the industry.

\subsection{Common Vulnerabilities and Exposures (CVE):}
The Common Vulnerabilities and Exposures (CVE) \cite{kronser2020common} initiative is a globally recognized program designed to identify and catalog cybersecurity vulnerabilities and exposures. Each CVE entry provides a standardized identifier, a brief description of the vulnerability, and references to related advisories or technical details, often including the affected codebase. CVEs play a critical role in vulnerability management by integrating with security tools like patching systems, enabling organizations to streamline mitigation efforts, protect their infrastructure, and enhance their security strategies.
\section{Approach}
\label{sec:approach}

\begin{figure*}
    \centering
    \includegraphics[width=0.98\textwidth]{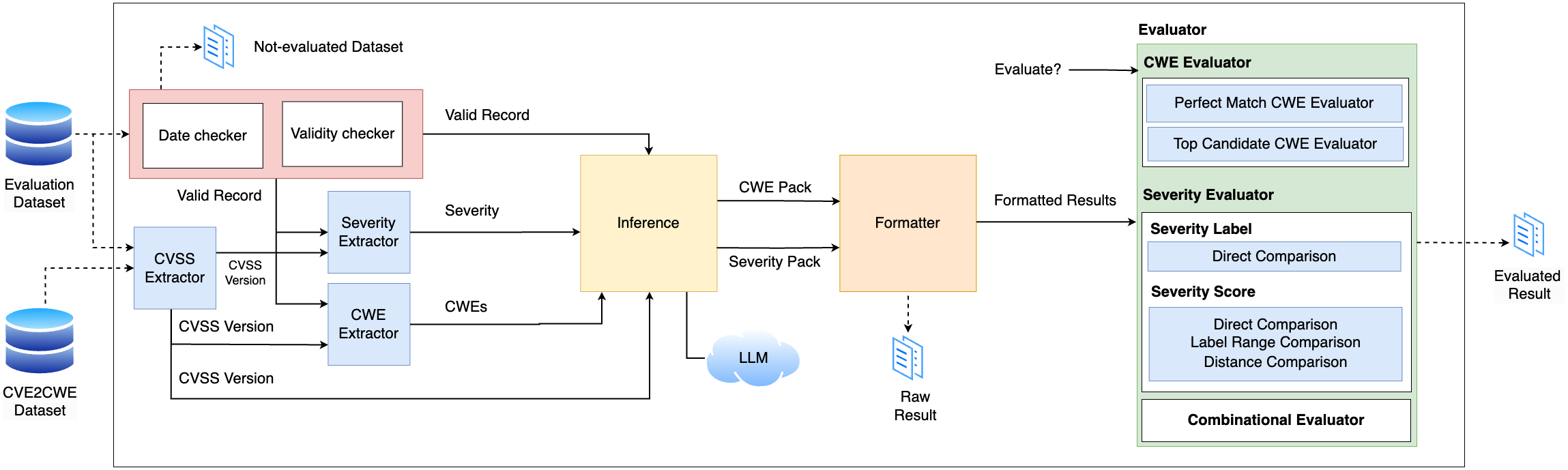}
    \caption{\textbf{\tool Workflow.}}
    \label{fig:workflow_diagram}
\end{figure*}

Our designed approach generates a reach prompt that is fed into a find-tuned language model to infer CWEs and severities and evaluate them accordingly. We have developed a prototype, \tool, as an end-to-end pipeline that requires two input datasets: one for evaluation and another for extracting CWEs and severities associated with given CVEs (called \texttt{CVE2CWE} dataset). It processes each record in the evaluation dataset, performs evaluations, and generates a comprehensive evaluation report as output. As shown in Figure \ref{fig:workflow_diagram}, \tool consists of several sequential components, where each component processes one or more inputs and provides outputs to the subsequent component(s). This section details the functionality of each component and describes how we fine-tuned the LLM to improve the inference accuracy.

First, each record (hereafter referred to as a candidate record) is fetched from the evaluation dataset. These CVE records are enhanced by augmenting buggy file contents and hunks. Given that the evaluation dataset may contain records with commit dates earlier than the LLM's training cutoff date, the record is first passed through the \textit{Date Checker} component. This component ensures the commit date is after the LLM's cutoff date, guaranteeing the record is unseen by the model. Records failing this condition are discarded into a report file named \texttt{Non-evaluated Dataset}. Candidate records passing the Date Checker are processed by the \textit{Validity Checker}, which verifies several conditions. These conditions include:
i) Non-empty, valid ground truth CWEs (as a list of CWE IDs) and severity information (as either CRITICAL, HIGH, MEDIUM, LOW), ii) Non-empty buggy code and hunks within the GitHub commit, iii) Supported programming languages for the committed files\footnote{Currently \tool supports nine programming languages, including C, C++, go, Java, JavaScript, Typescript, Ruby, Python, and PHP}, and iv) Compliance with the token size limit imposed by the GPT-3 model. Records failing these criteria are also discarded into the output file.

Next, using the \texttt{CVE2CWE} dataset (which holds the set of CVE IDs and their assigned CWEs and severities), the associated CVSS of the candidate record is extracted by the \textit{CVSS Extractor} component. This component takes the CVE (which has the ground truth for multiple CVSS versions) and the \texttt{CVE2CWE} dataset (which contains the CVSS version of all CVEs) and generates a list of CVSS versions for that specific CVE as output. Given that there could be multiple CVSS versions, \tool later chooses the latest CVSS for the given CVE, ensuring that the severity label and score are extracted according to the latest version announced by NIST. If the CVSS is unspecified for the given CVE, the default version (v3.1) is used. Given the CVSS version, the severity and score are retrieved from the corresponding record using the \textit{Severity Extractor} and \textit{CWE Extractor} components, respectively. These two components extract the severity and CWE fields (for that specific CVSS version) for that candidate record as their output. Therefore, this process enhances \tool’s consistency and effectiveness within different security assessment frameworks. It also helps streamline the bug triaging process and aids security teams in working with the most up-to-date vulnerability data.

Next, the record along with the extracted fields (CVSS version, severity score, and label) is passed through the \textit{Inference} component, where CWE and severity inferences are made. Depending on the experiment variant (as we discuss in Section \ref{sec:experiments}), it extracts the following elements from the candidate record depending on the experiment variant and makes inferences via the Application Programming Interface (API) of the language model:

\begin{itemize}
    \item \textbf{Bug description:} This field is required by all experiment variants and is directly extracted from the evaluation dataset record.
    \item \textbf{Buggy files:} If the variant requires the contents of buggy files, the content is directly extracted from the dataset and embedded in the prompt.
    \item \textbf{Buggy methods:} To extract the methods whose content was changed by the commit for bug resolution, \tool invokes the corresponding extractor scripts according to their programming language. These scripts extract the AST from the given buggy file and identify methods where the deleted buggy lines exist within the commit. This process is done using the \textit{tree\_sitter}\cite{tree_sitter} library.
    \item \textbf{Buggy hunks:} This field is also directly extracted from the evaluation dataset.
\end{itemize}

Then, the inference for CWE and severity (score and label) is conducted separately to ensure the inclusion of specific information in each prompt and to prevent overloading the LLM with combined information. After making two inferences, the output is forwarded to the \textit{Formatter} component to be syntactically checked and ensure that the output aligns with the expected format before being evaluated by the \textit{Evaluator} component. The metrics for this syntactic formatting include having a list of identified CWEs (one field for perfect match CWEs, and the other for top candidate CWEs), a numeric severity score (converted if originally provided as a string), and a string (or null) for the severity label. This component also ensures that the perfect match CWEs are a subset of the identified top candidate CWEs. The output is saved as a JSON file, provided that evaluation has been set in the tool using the \texttt{Evaluate} boolean flag. If this flag is not set, no evaluation will be performed. These output file makes the tool multi-stage, meaning that the results can be evaluated at any time by the \textit{Evaluator} component later without the need to run the tool again from scratch. This output file includes the LLM input and outputs (for CWE and severity), ground truth, CVSS version, any error messages encountered during the process, and the NVD description for the corresponding candidate record.

In the case of evaluation, \tool evaluates the JSON file against the corresponding ground truth for both perfect match and top candidate CWEs, as well as severity label and score. In this research, multiple combinations serve as criteria for assessing the accuracy of the inferences. These combinations are based on perfect matches (referred to as ``exact matches'' in the sample prompt) and top candidate CWEs (referred to as ``top-5 CWEs'' in the sample prompt), as well as the severity label and score. For CWE comparison, the inferred and ground truth sets are checked against equality and subset equality relationships. In view of the severity labels, we evaluate label equivalence between the ground truth label and the identified label. Regarding the severity score, three measurement criteria are considered:

\begin{enumerate}
    \item Direct comparison: Checking if the ground truth severity score is equal to the inferred severity score.
    \item Label range comparison: Verifying if both the ground truth and inferred severity scores fall within the same severity label range. For instance, in CVSS v3.1, scores of 5 and 5.4 both correspond to the MEDIUM severity label.
    \item Score distance comparison: Evaluating whether the ground truth severity score lies within a defined range \([score-distance, score+distance]\), with distances including 0.5, 1.0, and 1.5 considered.
\end{enumerate}

The \textit{Evaluator} component assesses each raw result against the ground truth and assigns a corresponding status for both CWE and severity. These status values include equal, subset-equal, overlapped, and non-overlapped for CWE, as well as the corresponding severity score status for each of the three score comparison criteria. Additionally, this component computes the tool's accuracy for each combination of CWE and severity (label and score), as detailed in Section \ref{sec:experiments}.

\subsection{Fine-tuning LLM}
Given that the standard model is used as a baseline, we fine-tuned the GPT-3 model to evaluate the accuracy of our tool within the proposed approach. This fine-tuning was performed on two unseen datasets (distinct from the evaluation dataset): one for CWE prediction and the other for severity. Each tuning dataset was converted into a map of prompts and their associated CWEs or severities. To create each fine-tuning record in either dataset, a corresponding prompt was created and tailored to include the required information (e.g., a description or a description with code), asking the model to predict CWEs or severity, depending on the inference. This process, known as prompt engineering, ensures that if a variant requires the inclusion of buggy code, the code (at the required granularity level) is included in the prompt. Note that for fine-tuning, all variants intended for evaluation were considered. This means the model is fine-tuned on all granularity levels for both CWE and severity identification so that it can used by the tool for each variant without needing to have a distinct model for each variant. Fine-tuning was performed with default metrics suggested by the language model's provider company using APIs. Table \ref{tab:fine_tuning_details} provides the configuration fine-tuning details and specific adjustments. As discussed earlier, the CWE fine-tuned model focuses on identifying common weaknesses in buggy code, while adopting the severity model by the tool results in identifying the severity.

\begin{table}
    \caption{Fine-tuning Parameters and Metrics.}
    \label{tab:fine_tuning_details}
    \centering
    \renewcommand{\arraystretch}{1.6}
    \setlength{\tabcolsep}{10pt}
    \begin{tabular}{|l|c|c|}
        \hline
        \textbf{Metric} & \textbf{CWE Model} & \textbf{Severity Model} \\
        \hline
        \textbf{Training Tokens} & 46,492,872 & 54,682,551 \\
        \hline
        \textbf{Epochs} & Default (3) & Default (3) \\
        \hline
        \textbf{Batch Size} & Default (11) & Default (11) \\
        \hline
        \textbf{LR Multiplier} & Default (2) & Default (2) \\
        \hline
        \textbf{Training Loss} & 0.0318 & 0.0346 \\
        \hline
        \textbf{Validation Loss} & 0.0789 & 0.1241 \\
        \hline
    \end{tabular}
\end{table}

\begin{figure}[h]
\centering
\includegraphics[width=0.48\textwidth,height=0.42\textheight]{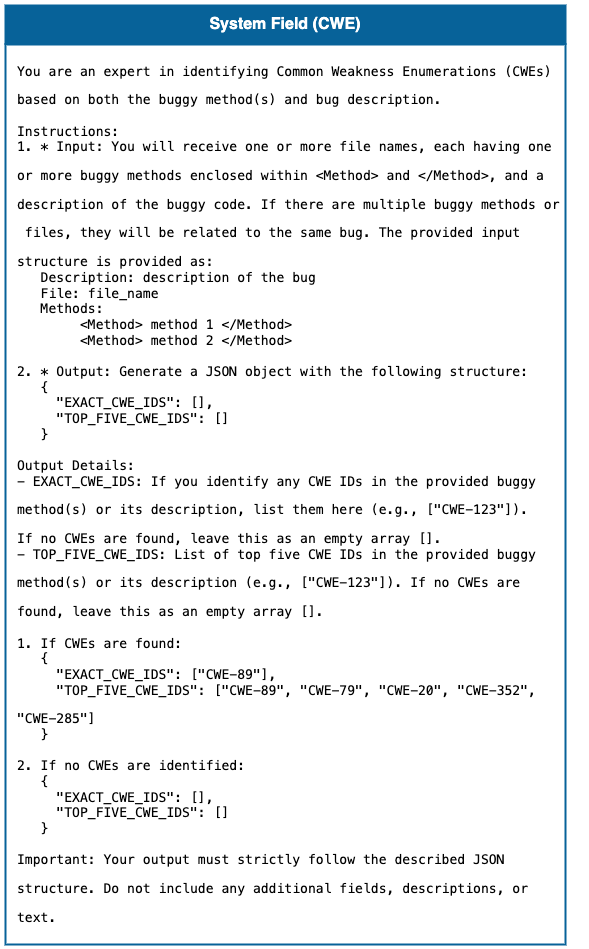}
\caption{\label{fig:sample_of_system_field_cwe}Prompt Sample (System Field - CWE).}
\end{figure}

\begin{figure}[h]
\centering
\includegraphics[width=0.48\textwidth, height=0.35\textheight]{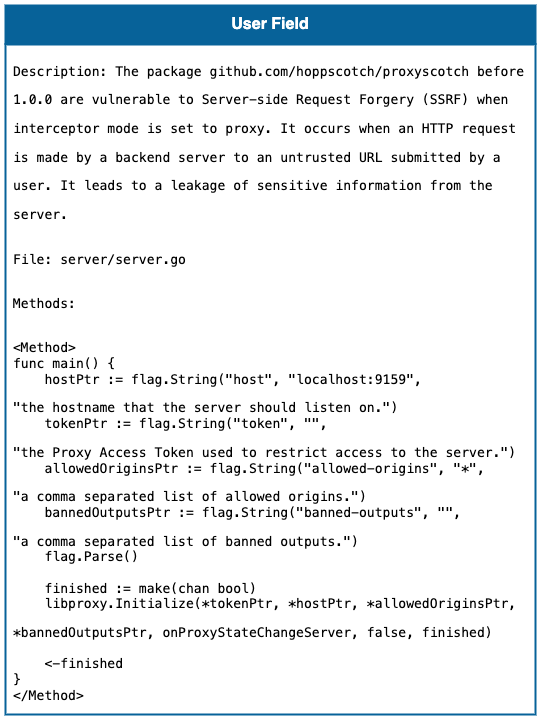}
\caption{\label{fig:user_field}Prompt Sample (User Field).}
\label{fig:sample_user_field}
\end{figure}

\begin{figure}[h]
\centering
\includegraphics[width=0.48\textwidth]{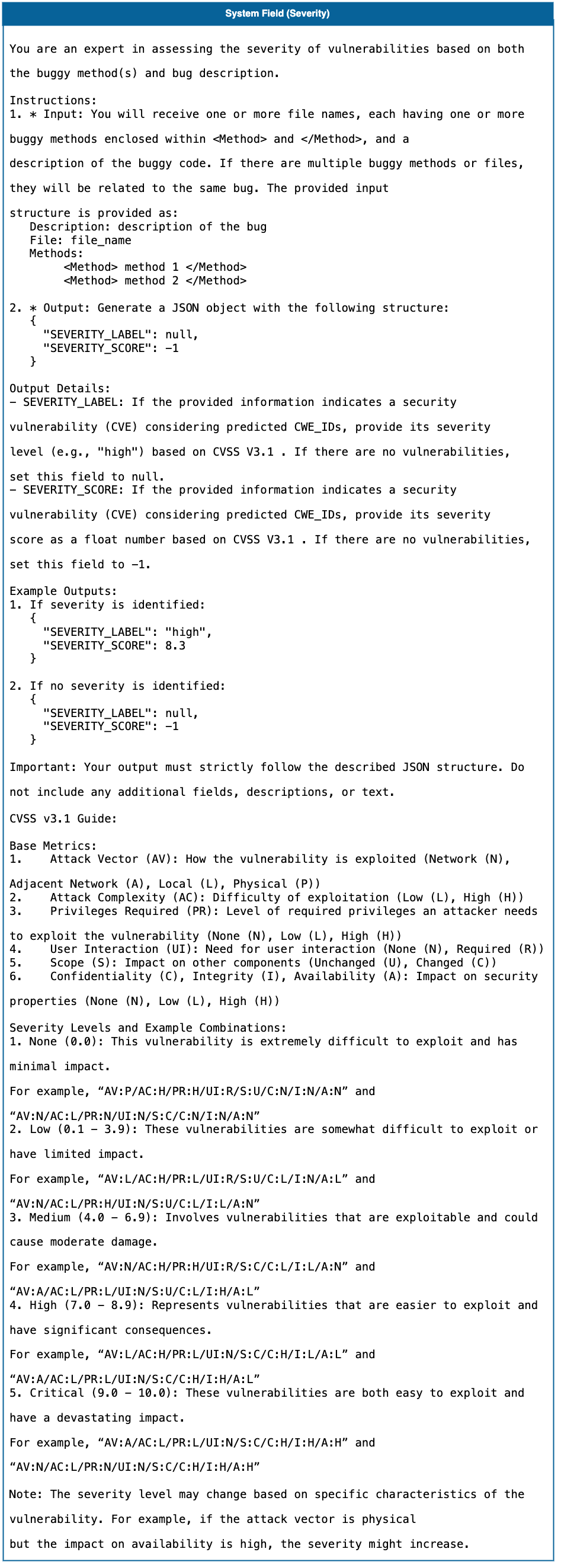}
\caption{\label{fig:example_of_system_field_severity}Prompt Sample (System Field - Severity).}
\end{figure}

Appendix \ref{sec:appendix_a} provides samples of the \textit{system} and \textit{user} fields (Figures \ref{fig:sample_of_system_field_cwe}, \ref{fig:sample_user_field}, and \ref{fig:example_of_system_field_severity}) used by the tool for the ``description + methods'' variant. Overall, the \textit{system} prompt (Figure \ref{fig:example_of_system_field_severity}) provides the model with the following information:

\begin{itemize}
     \item Persona adoption: The model is instructed to adopt a specific persona. In this case, an expert in severity identification.
    \item Input descriptions: The model is informed about the expected inputs, along with the template format in which the \textit{user} field provides information.
    \item Output descriptions: The model is given details about the expected inference output, along with the template format it should use to generate the output.
   \item Default value descriptions: If the model cannot identify any severity for the given vulnerability, default output values are explicitly described (\texttt{null} for severity label and -1 for severity score).
    \item CVSS descriptions: Based on the input record's CVSS version (extracted from the CVE record of the evaluation dataset), the model is provided with a brief guide on CVSS, including attack vector examples for each severity label.
\end{itemize}

This information is specifically tailored for the severity inference \textit{system} prompt, where a CVSS description is provided to the model. For the CWE inference (Figure \ref{fig:sample_of_system_field_cwe}) and also for the other experiment variants, the instructions—particularly the inputs provided to the model—are customized to guide the model effectively during inference. For the \textit{user} field, the model receives record-specific information depending on the experiment variant. Figure \ref{fig:sample_user_field} illustrates the information provided for the ``description + buggy method'' variant, where the bug description and buggy method, along with the container file name, are enclosed within \verb|<Method></Method>| tags to enhance precision during inference. The prompts for each experimental record across all variants are included in the released artifacts.

With the suggested approach, \tool offers two key use cases: 
\begin{enumerate}
    \item \textbf{New bug CWE and severity identification:} By analyzing the NVD dataset, we found 793 records (i.e., CVEs) out of 6,032 records where either the CWE or severity has not been determined, despite having non-empty associated descriptions. These incomplete CVEs pose challenges for developers relying on them. In such scenarios, where a new bug's (CVE's) associated CWE or severity is under review, \tool can be used to efficiently determine these fields. This allows developers to classify newly discovered bugs by identifying their CWEs and assigning appropriate severities based on the most recent version of CVSS that \tool is expert. Moreover, given that the tool is compatible with any non-NVD datasets (including the vulnerabilities found by developers) as long as the required inputs are provided to the tool, it can be integrated with any static analyzer or fuzzers to identify the CWE and severity score, which enables product teams to prioritize vulnerabilities. We tested our approach on such records and explained the evaluation results in Section \ref{sec:results}. We should note that in real-world scenarios, it is quite common for automated bug-finding approaches, such as fuzzing \cite{8418632, 8371326, 10.1145/3512345}, to result in numerous crashes where the locations of the causing bugs are known. However, developers may not always have insights into their associated CWEs or severities. In such scenarios, \tool can assist developers and enterprises in prioritizing crashes and identifying their types.
    
    \item \textbf{Postmortem identification:} \tool can also help developers efficiently manage previously triaged bug vulnerabilities, ensuring historical data remains aligned with evolving CVSS standards. This involves re-evaluating existing bug reports to keep historical vulnerability data consistent with updated CVSS versions. As the CVSS scoring system evolves, older vulnerabilities may need reassessment to reflect updated severity metrics. \tool streamlines this process by analyzing the associated buggy code (e.g., code snippets, buggy methods, or buggy hunks) and descriptions to update CWEs and CVSS severity levels in accordance with the latest standards. For example, the CVE-2014-0160, known as the Heartbleed vulnerability, had its severity labeled as MEDIUM under CVSS 2.0 and was classified as HIGH in CVSS 3.x. This inconsistency incorrectly equates Heartbleed's severity with less severe CVEs, even though it is widely recognized as one of the most critical vulnerabilities, allowing attackers to read memory without restriction. Ideally, Heartbleed should have a severity of CRITICAL under modern CVSS standards. \tool can assist in such scenarios by re-identifying the severities of previously identified bugs based on updated CVSS knowledge. When new CVSS versions are released, \tool enables security teams to reassess older vulnerabilities efficiently, ensuring a more accurate understanding of their organization's risk landscape. 
\end{enumerate}

\section{Implementation}
\label{sec:implementation}

The pipeline has been implemented in Python. We utilized the \textit{tree\_sitter} library \cite{tree_sitter} to extract methods from a given buggy file. The pipeline comprises two stages: experimentation and evaluation, enabled by the \texttt{Evaluate} flag, as shown in Figure \ref{fig:workflow_diagram}. The experimentation stage generates raw JSON results containing inferences of CWEs, severity levels, and the corresponding ground truth data, while the evaluation stage analyzes these raw results against ground truth and produces the evaluation results.

Regarding the fine-tuning, as depicted in Table \ref{tab:fine_tuning_details}, the training loss for both models is nearly identical. However, the CWE model has a lower validation loss, indicating slightly better accuracy in identifying CWEs for the validation dataset. The severity model required more training tokens than the CWE model, due to the verbosity of their system fields, as shown in Figures \ref{fig:sample_of_system_field_cwe} and \ref{fig:example_of_system_field_severity} in Appendix \ref{sec:appendix_a}. The LLM follows a set of \textit{system} instructions that guide its inferences, and the \textit{user} field that provides instance-specific information tailored to each candidate record (Figure \ref{fig:user_field} in Appendix \ref{sec:appendix_a}). The GPT-3 model's cut-off date is considered to be 2021-09-01 so the datasets' records (both evaluation and fine-tuning datasets) that are after this date are discarded by the tool.
\section{Experiments}
\label{sec:experiments}

In this section, we explain how we collect the data required to address each research question. We also describe the terms and metrics used to measure the results, the design of the experiments to answer each research question, as well as the process of fine-tuning the LLM model.

\subsection{Data Collection}

\begin{figure*}
    \centering
    \includegraphics[width=0.7\textwidth]{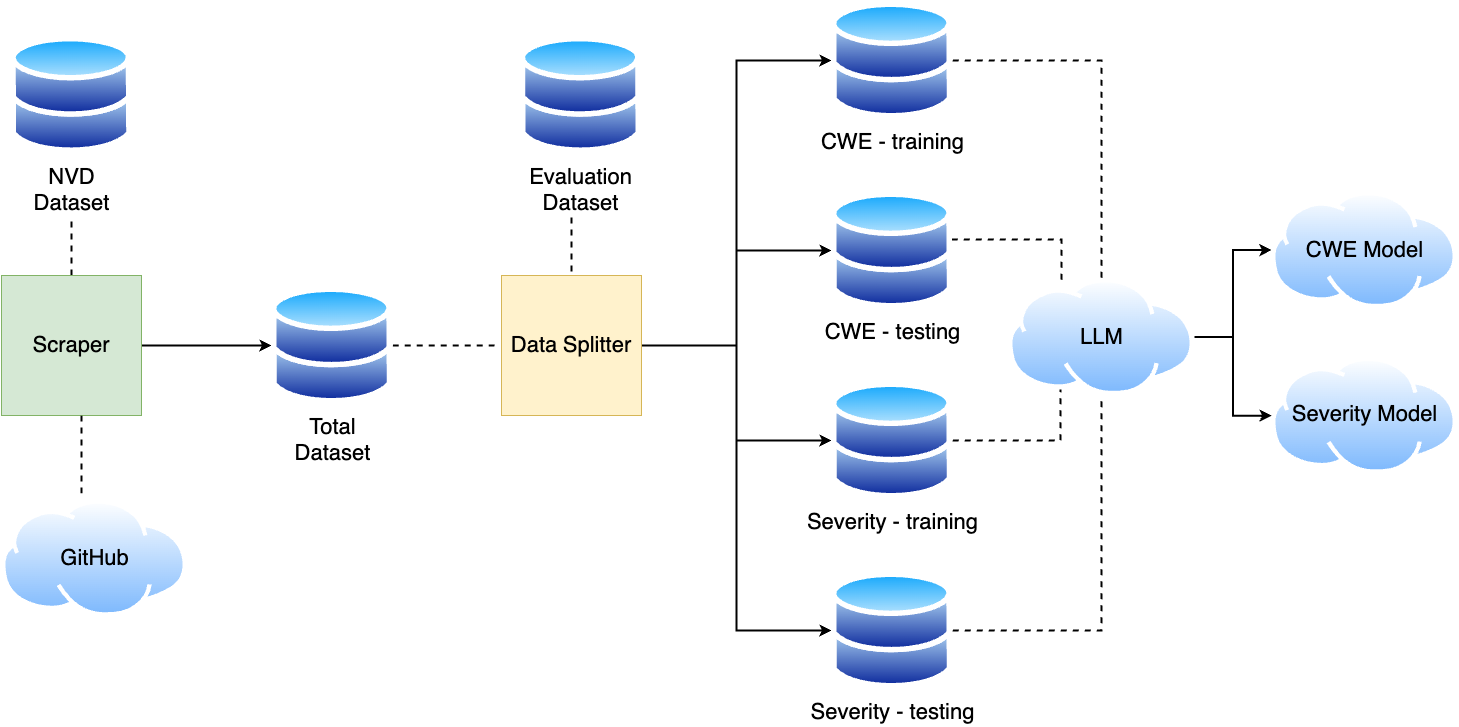}
    \caption{\textbf{Dataset Creation Workflow.}}
    \label{fig:dataset_creation_diagram}
\end{figure*}

Figure \ref{fig:dataset_creation_diagram} illustrates the process our approach performs to collect the evaluation dataset as well as the fine-tuning datasets required to fine-tune the models for CWE and severity inferences. We use the National Vulnerability Database (NVD) as our primary data source. It provides various information about security vulnerabilities, including associated CWEs, CVSS scores, and links to GitHub repositories containing relevant bug reports and code. We enhance this JSON dataset by inserting the corresponding buggy codes for each record fetched from their associated GitHub commit URL. We scrape the buggy files from their GitHub commits and embed them in the repository to form the final dataset. To scrape the buggy code, we first cloned each project's repository at the specific commit version. We then incorporated the buggy code at each granularity level (file, method, and hunk) into our dataset. This dataset includes the following items:

\begin{enumerate}
    \item \textbf{Buggy code (array):} An array of JSON objects, each containing the buggy file name, buggy file content, and buggy lines along with their line numbers. We developed a script to scrape the deleted lines in the commit URL, capturing these as buggy lines resolved after the commit.
    \item \textbf{Description (string):} The template-based NVD bug description of the CVE record.
    \item \textbf{URL (string):} The GitHub commit URL associated with the CVE.
    \item \textbf{CVE (string):} The identifier of the CVE record.
    \item \textbf{Date (string):} The date when the commit was pushed.
    \item \textbf{GitHub description (string):} If the commit is linked to an issue, this field contains the issue message. Otherwise, it is set to \texttt{null}.
\end{enumerate}

Next, we split this dataset (consisting of 6,032 candidate records) into two distinct subsets: one for evaluation and the other for fine-tuning. This separation ensures the datasets remain independent, maintaining the fairness and validity of the evaluation by preventing exposure to fine-tuning records during the evaluation process. Table \ref{tab:dataset_statistics} provides details on the size of each subset.

\begin{table}[h]
    \centering
    \normalsize
    \caption{Dataset Statistics}
    \begin{tabular}{|l|c|}
        \hline
        \textbf{Datasets} & \textbf{Record Count} \\
        \hline
        Initial Dataset & 3,016 \\
        Filtered Dataset & 1,155 \\
        Prepared For Seven Experiments & 8,085 \\
        Token Filter Applied (CWE) & 7,933 \\
        Token Filter Applied (Severity) & 7,894 \\
        \hline
        Training Dataset (CWE) & 5,949 \\
        Testing Dataset (CWE) & 1,984 \\
        Training Dataset (Severity) & 5,920 \\
        Testing Dataset (Severity) & 1,974 \\
        \hline
        Evaluation Dataset (Total Records) & 3,016 \\
        Selected Evaluation Records & 500 \\
        \hline
    \end{tabular}
    \label{tab:dataset_statistics}
\end{table}

As presented in Table \ref{tab:dataset_statistics}, the dataset used in this study underwent multiple stages of preparation to ensure its suitability for evaluation and fine-tuning. That means that for the evaluation process, a separate evaluation dataset comprising 3,016 records (i.e., 50\%) was split, entirely distinct from the fine-tuning dataset. From this evaluation dataset, 500 records were randomly selected for evaluation (for each record, 7 records were generated for each experiment variant). For fine-tuning, the initial dataset consisted of 3,016 records. After applying two filtering processes (by \textit{Date Checker} and \textit{Validity Checker} components) as described in Section \ref{sec:approach}, 1,155 records remained, forming the foundation for generating additional records through seven experimental procedures. For each record, a separate record was generated containing the required information (including the buggy code with the appropriate granularity level) for that experiment variant. This process resulted in a total of 1,155*7=8,085 records. To further refine the dataset, a token filter was applied to ensure that our final prompt adheres to the token number limits of the language model, reducing the number of records to 7,933 for the CWE identification task and 7,894 for the severity identification task. The dataset was then divided into training and testing sets. For the CWE task, 5,949 records (75\%) were allocated for training and 1,984 records (25\%) for testing. Similarly, for the severity task, 5,920 records (75\%) were used for training, and 1,974 records (25\%) for testing.

In terms of dataset distribution, Table \ref{tab:severity_freq_dist} presents the distribution of severity labels (HIGH, MEDIUM, and LOW) across the total, evaluation, and fine-tuning datasets. MEDIUM severity instances are the most prevalent in all datasets, followed by HIGH and LOW severity instances. This overrepresentation of MEDIUM severity is due to the random selection of records when splitting the total dataset, which led to a higher proportion of MEDIUM instances in both the total and fine-tuning datasets.

\begin{table}
\centering
\caption{Severity Label Distribution}
\renewcommand{\arraystretch}{1.5} 
\begin{tabular}{|l|c|c|c|}
\hline
\textbf{Dataset} & HIGH & MEDIUM & LOW \\
\hline
\textbf{Total} & 241 & 1,247 & 345 \\
\textbf{Evaluation} & 113 & 646 & 179 \\
\textbf{Fine-tuning} & 128 & 601 & 166 \\
\hline
\end{tabular}
\label{tab:severity_freq_dist}
\end{table}

In view of CWE distribution across datasets, Table \ref{tab:cwe_freq_dist} shows the distribution of CWE categories across four ranges: [1-500], [501-1000], [1001-1336], and Unknown. The majority of instances across all datasets fall within the [1-500] range, indicating a higher prevalence of lower-numbered CWE categories which often indicate the fundamental code weaknesses. This trend is consistent across the total, evaluation, and fine-tuning datasets, where instances in the [1-500] range dominate the distribution. In contrast, the [1001-1336] range and the Unknown category represent a significantly smaller portion of the datasets. This distribution suggests that lower-numbered CWEs are more common, while higher-numbered and unknown CWEs are less frequent.

\begin{table}
\centering
\caption{CWE Frequency Distribution}
\renewcommand{\arraystretch}{1.5}
\begin{tabular}{|l|c|c|c|c|}
\hline
\textbf{Dataset} & [1-500] & [501-1,000] & [1,001-1,336] & UNK. \\
\hline
\textbf{Total} & 3852 & 1,271 & 170 & 118 \\
\textbf{Evaluation} & 1,883 & 644 & 84 & 67 \\
\textbf{Fine-tuning} & 1,969 & 627 & 86 & 51 \\
\hline
\end{tabular}
\label{tab:cwe_freq_dist}
\end{table}

Table \ref{tab:dataset_pr_language_statistics} illustrates the distribution of programming languages across the evaluation, fine-tuning, and total datasets. The most frequent languages across all datasets are PHP, JavaScript, and TypeScript or Ruby, while languages such as Scala and Swift are the least represented. This trend indicates a dominance of certain widely-used programming languages, particularly PHP and JavaScript, while less common languages like Scala and Swift appear infrequently across the datasets.

\begin{table}
\caption{Programming Languages Distribution}
\centering
\resizebox{0.46\textwidth}{!}{
\renewcommand{\arraystretch}{1.5} 
\begin{tabular}{|l|l|c|}
\hline
\textbf{Dataset} & \textbf{Most Occurred Languages} & \textbf{Least Occurred Languages} \\
\hline
\textbf{Evaluation} & php: 5,689, js: 1,237, ts: 951 & scala: 53, swift: 44 \\
\hline
\textbf{Fine-tuning} & php: 4,456, js: 1,354, rb: 1,235 & swift: 25, scala: 10 \\
\hline
\textbf{Total} & php: 10,145, js: 2,591, py: 1,949 & swift: 69, scala: 63 \\
\hline
\end{tabular}
}
\label{tab:dataset_pr_language_statistics}
\end{table}

\subsection{Experiments Design}
\subsubsection{Baseline Experiments}
For the baseline experiments, we conducted experiments by making the tool use the standard GPT-3 model. To determine the experiment variants, we conducted small-scale experiments on a limited dataset (200 samples). These experiments revealed that providing the model with buggy files, buggy methods, or buggy hunks resulted in low accuracy (less than 10\%). However, supplying the model with the bug's NVD description or its combination with buggy code produced more promising results. Consequently, due to resource constraints for such variants, we exclude them and select only the four mentioned variants (prompts):  1) Description, 2) Description + (buggy) files, 3) Description + (buggy) methods, 4) Description + (buggy) hunks. Moreover, this shows that adding a description to the buggy code boosts the \tool's accuracy in identifying both CWE and severity. Another note is that providing the tool with a bug description is recommended for newly identified bugs, where no buggy code has been identified at any granularity. On the other hand, when updating the CWEs or severities of CVEs (post-mortem analysis), other variants (prompts) can be used to refine the classification or generate updated severity labels, ensuring the tool remains effective and adaptable for evolving bug reports.

For each experiment variant, we performed two separate inferences with different system-user field pairs. For CWE identification, we asked the model to provide the perfect match and top (five) candidate CWEs related to the given candidate record. For severity identification, we asked the model to provide both the severity label and severity score.

\subsection{Evaluation Metrics}
For CWE identification, we measure the accuracy in terms of generating CWEs (both perfect match and top candidates) that are equal to the ground truth, calling this ``Prediction Equality (PE)''. This metric is computed as the proportion of the candidate records holding this condition and is computed as shown below:

\[
PE Accuracy =
    \frac{\#Identified Records Having Equal CWEs}{Total Number of Records}
\]

In terms of identifying whether the set of identified CWEs match or are smaller/larger than the ground truth, we define two metrics:

\begin{enumerate}
    \item Prediction Coverage (PC): This metric captures when the identified CWEs are a subset of (or equal to) the ground truth. In other words, the identified set of CWEs is smaller than or equal to the ground truth CWEs.
    \item Ground truth Coverage (GC): This metric captures when the ground truth CWEs are a subset of (or equal to) the identified CWEs. In other words, the identified set of CWEs is larger than or equal to the ground truth CWEs.

\end{enumerate}

We calculate these two metrics as follows:

\[
PC Accuracy =
    \frac{\#Identified Records Having Smaller CWEs}{Total Number of Records}
\]

\[
GC Accuracy =
    \frac{\#Identified Records Having Larger CWEs}{Total Number of Records}
\]

For severity, we have divided our metrics into two main categories and computed the accuracy as the proportion of the records (compared to the entire evaluation set size) having each property.

\begin{enumerate}
    \item \textbf{Severity Label:} To evaluate the correctness of the severity label, we directly compare the ground truth severity label with the inferred severity label (case-sensitively) and divide its frequency by the total number of ground truth records. There are four possible labels for the severity label inferences: LOW, MEDIUM, HIGH, and CRITICAL.
    \item \textbf{Severity Score:} For this metric, we consider different criteria as follows:
    \begin{itemize}
        \item \textbf{Score Comparison:} This criterion evaluates the correctness of the severity scores by directly comparing the inferred severity score (as a floating-point number) with the ground truth. The severity score ranges from 0 to 10, covering three or four severity labels depending on the associated CVSS\footnote{The CVSS v2.0 has three severity labels within this range, while newer versions have four labels.}. Then we compute the proportion of the correctly identified severity scores, over the total number of ground truth records.
        \item \textbf{Label Range:} This criterion evaluates the equality of severity scores based on their associated label range. For an inferred severity score (e.g., 9.2) with a specific CVSS version (e.g., CVSS v3.1), the enclosing range for that score is extracted (in this case, [9.0-10.0]) and is checked against the ground truth severity label (in this case, CRITICAL). Finally, we compute the proportion of cases where the enclosing label range matches the ground truth severity label, over the total number of ground truth records. This approach mitigates the effect of severity score comparison inaccuracies by considering the associated label rather than the numerical comparison, aiming for greater accuracy.
        \item \textbf{Distance:} This criterion evaluates the severity scores by considering three different distances (0.5, 1.0, and 1.5). For each distance, a range is created around the inferred severity score by adding the distance to the score for the upper bound and subtracting it for the lower bound. This range is then checked against the ground truth severity score to see if it covers. For instance, considering a distance of 1.5 for an inferred severity score of 9.2 (with the ground truth being 8), the range would be adjusted to the minimum and maximum allowed bounds. In this case, the generated range would be [7.7-10.0] instead of [7.7-10.7]. This range covers the ground truth severity score. This metric is independent of severity labels and allows for a margin of error when evaluating if the generated range covers the ground truth severity score. We compute the proportion of cases where the range covers the ground truth severity score, over the total number of ground truth records.
    \end{itemize}
\end{enumerate}

\textbf{Total Accuracy:} It is evaluated by considering a perfect match between predicted and ground truth CWEs, along with severity labels and severity scores assessed based on label range and score distance. For each combination of these criteria, we define the overall accuracy formula as follows:

\[
Total Accuracy =
    \frac{\#Records Satisfying The Conditions}{Total Number of Records}
\]

Here, the conditions correspond to one of the specified combinations of criteria as i) perfect match CWE + severity label. ii) perfect match CWE + severity score (label range). iii) perfect match CWE + severity score (score distance)).
\section{Results}
\label{sec:results}

In this section, we explain the results of our experiments (compared to the baseline) with diagrams created for each criterion and answer each research question accordingly.

\textbf{RQ1: Can \tool accurately identify CWEs?}: Overall, our experiments demonstrate that integrating the tool with our fine-tuned model significantly improves the accuracy in identifying CWEs compared to the baseline. To answer this question, we will analyze both perfect matches and top candidate CWEs when adopting the fine-tuned models compared to the baseline experiments. \footnote{The x-axis represents the prompt we use for that experiment variant}:

\begin{figure}[t]
\centering
\includegraphics[width=0.48\textwidth]{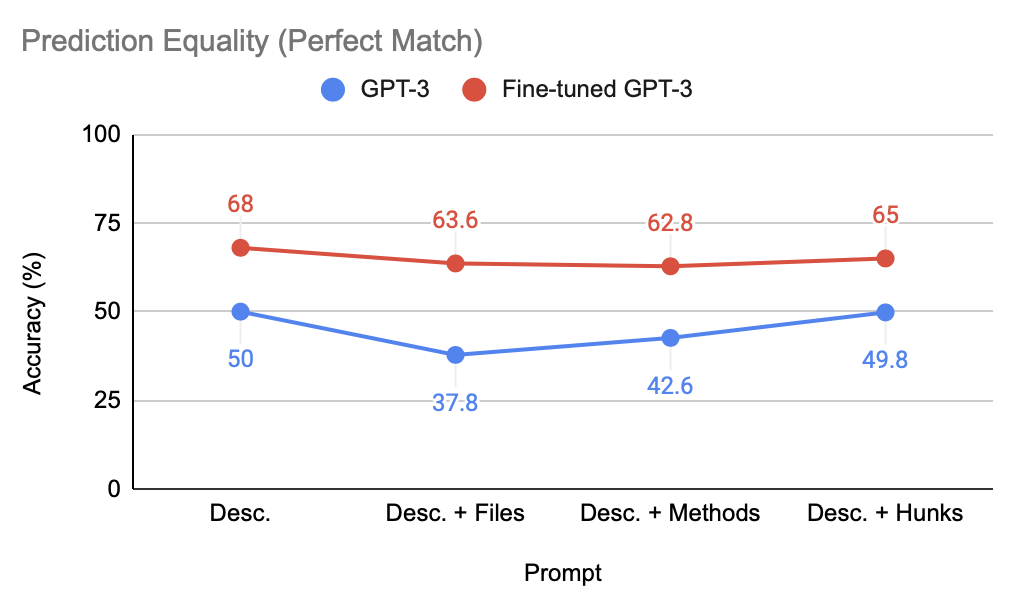}
\caption{\label{fig:gt_equal_e}Prediction Equality Accuracy (Perfect Match).}
\end{figure}

\begin{figure}[t]
\centering
\includegraphics[width=0.48\textwidth]{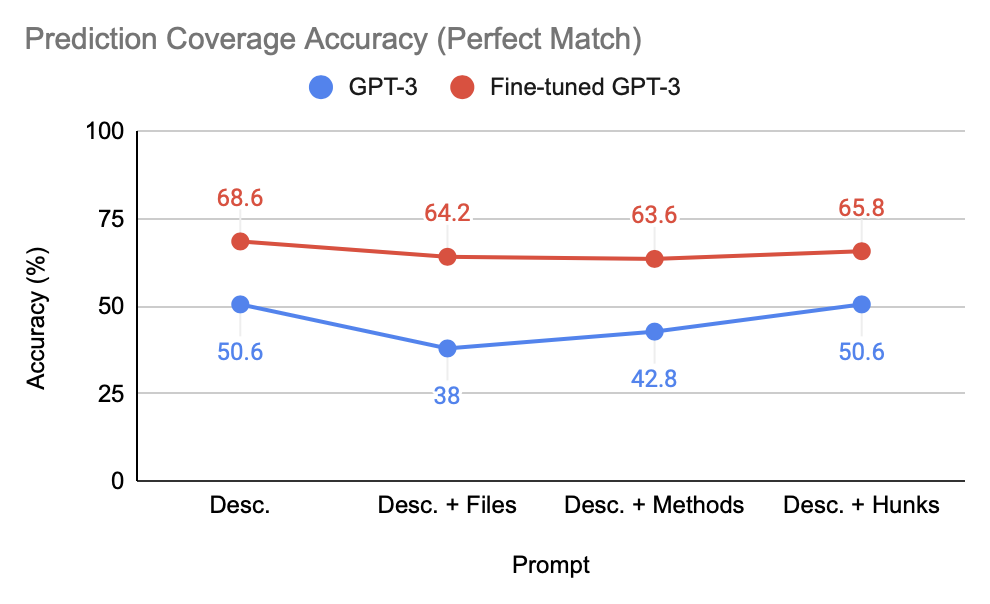}
\caption{\label{fig:e_subsetequal_gt}Prediction Coverage Accuracy (Perfect Match $\subseteq$ Ground Truth).}
\end{figure}

\begin{figure}[t]
\centering
\includegraphics[width=0.48\textwidth]{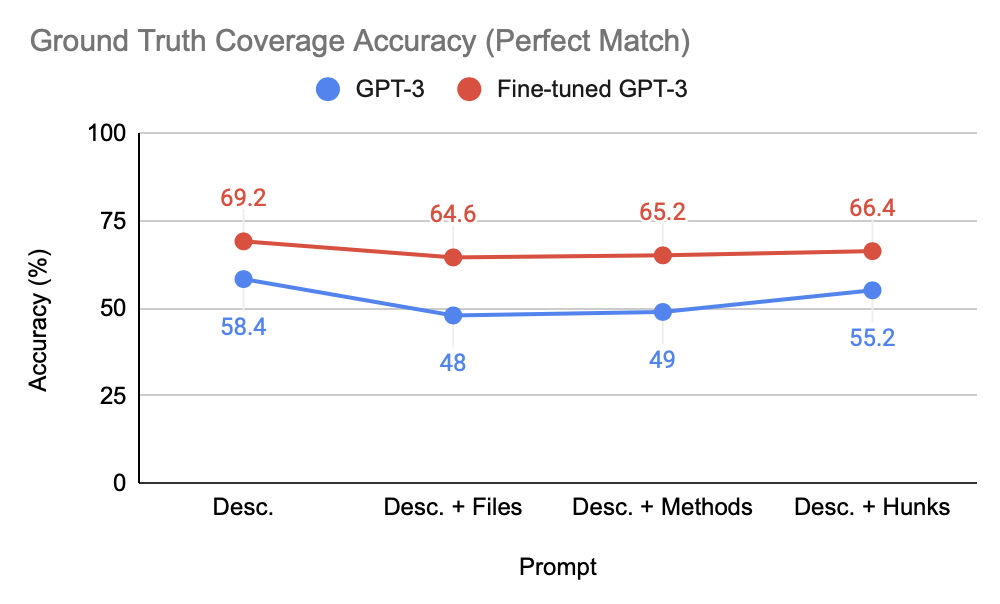}
\caption{\label{fig:gt_subsetequal_e}Ground Truth Coverage Accuracy (Ground Truth $\subseteq$ Perfect Match).}
\end{figure}

\textbf{Perfect Match CWEs:} Figure \ref{fig:gt_equal_e} illustrates the accuracy of \tool in identifying perfect match CWEs that are equal to the ground truth CWEs. The accuracy ranges from 62.8\% to 68.0\% when using the fine-tuned model (red), compared to the baseline (blue), which ranges from 37.8\% to 50.0\%. This represents an average improvement of 19.8\% over the baseline across all prompts. While \tool's performance with the fine-tuned model remains relatively stable, there is a maximum difference of 12.2\% between the lowest accuracy (achieved with the ``description + file'' prompt) and the highest accuracy (achieved with the ``description'' prompt). For the baseline, providing the code at the file granularity level along with the description results in the lowest accuracy, whereas replacing the file-level granularity with finer granularity (methods and hunks) leads to gradual improvements in accuracy. In contrast, for the fine-tuned model, accuracy across different granularity levels remains almost stable. However, similar to the baseline, providing hunks along with the description yields the highest accuracy. A notable observation from this analysis is that providing only the description better guides the model toward identifying perfect match CWEs, whereas including code alongside the description lowers accuracy, being a distractor. Additionally, in the baseline, providing the buggy file overwhelms \tool with excessive information, which negatively impacts its ability to identify CWEs. This pattern is evident across all prompts in Figures \ref{fig:gt_equal_e}, \ref{fig:e_subsetequal_gt}, and \ref{fig:gt_subsetequal_e}.

To assess whether the ground truth CWEs cover the identified perfect match CWEs (or vice versa), Figure \ref{fig:e_subsetequal_gt} shows the ``prediction coverage'' since it shows the accuracy of \tool in terms of identifying perfect match CWEs that are a subset or equal to the ground truth CWEs. The figure shows that the accuracy of the tool for this metric ranges from 63.6\% to 68.6\%, compared to 38\% to 50.6\% in the baseline. This shows the superiority of our approach in terms of identifying perfect match CWEs that are a subset or equal to the ground truth CWE. For the reverse direction (referred to as ``ground truth coverage''), Figure \ref{fig:gt_subsetequal_e} describes the accuracy when the ground truth is a subset or equal to the identified perfect match CWEs. In this figure, the accuracy of the tool ranges from 64.6\% to 69.2\%, standing above the baseline experiment (which ranges from 48\% to 58.4\%), showing the effectiveness of the contribution. Putting Figures \ref{fig:gt_equal_e}, \ref{fig:e_subsetequal_gt}, and \ref{fig:gt_subsetequal_e} together, it is concluded that while our approach generated more equal perfect match CWEs rather than generating CWEs that are subset/superset of the ground truth CWEs, it overpredicts the perfect match CWEs, meaning that it identifies additional perfect match CWEs.

\begin{figure}[t]
\centering
\includegraphics[width=0.5\textwidth]{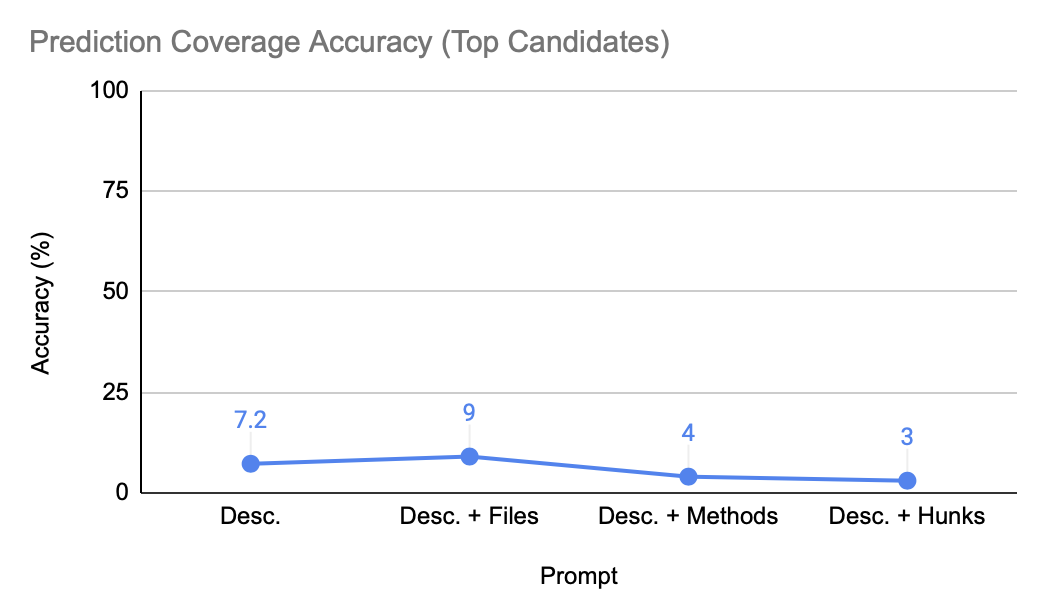}
\caption{\label{fig:t_subsetequal_gt}Prediction Coverage Accuracy (Top-candidates $\subseteq$ Ground Truth).}
\end{figure}

\begin{figure}[t]
\centering
\includegraphics[width=0.5\textwidth]{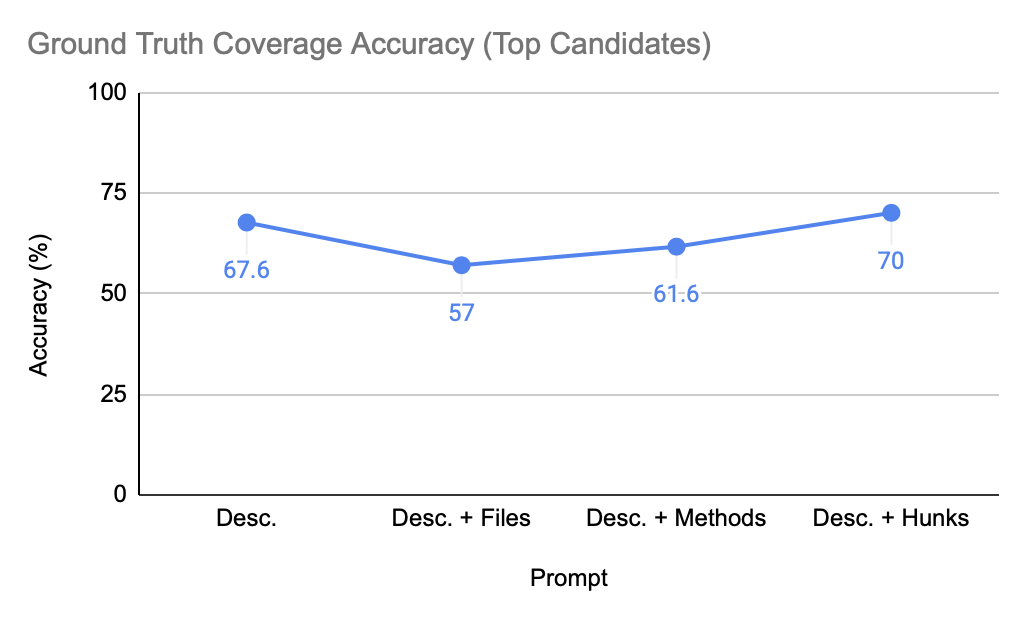}
\caption{\label{fig:gt_subsetequal_t}Ground Truth Coverage Accuracy (Ground Truth $\subseteq$ Top Candidates).}
\end{figure}

\textbf{Top Candidate CWEs:} In terms of the top five candidate CWEs in baseline, our findings show that the tool overpredicted in baseline, meaning that generated more cases in which the top candidate CWEs are either equal or superset of the ground truth CWEs. This claim is supported by comparing the Figures \ref{fig:t_subsetequal_gt} and \ref{fig:gt_subsetequal_t}, showing the reliability of the top candidate CWEs in terms of covering the expected CWEs. While this overprediction can occasionally provide valuable context by identifying related vulnerabilities, it may also introduce unnecessary noise to the identified CWEs. Regarding prompt comparisons in Figure \ref{fig:gt_subsetequal_t}, the diagrams show the effectiveness of providing buggy hunks along with the description compared to all other prompts. This tendency suggests that the tool, when using the standard model, generates a broader set of predictions, including additional CWEs beyond the ground truth. As depicted in Figure \ref{fig:gt_subsetequal_t}, the accuracy ranges from 57\% to 70\% in terms of over-prediction, with a minimum accuracy of 57\% in the ``description + file'' (which is, again, related to the excessive information) and a maximum accuracy of 70\% in the ``description + hunk'' prompt, which is related to a finer granularity. In addition to the discussed insights, a comparison of Figures \ref{fig:gt_subsetequal_e} and \ref{fig:gt_subsetequal_t} reveals that identifying the top candidate CWEs improves coverage of the ground truth compared to perfect matches. However, this comes at the cost of including CWEs that may have lower relevance to the actual ground truth, presenting a tradeoff between coverage and precision.

Note that given that no top candidate CWEs were used during the fine-tuning of the model, to ensure a fair comparison for \tool when adopting the two LLMs, we did not consider the tool's accuracy for the fine-tuned model in terms of top candidates (Figures \ref{fig:t_subsetequal_gt} and \ref{fig:gt_subsetequal_t}).

\begin{table}[h]
    \label{box:rq1_summary}
    \begin{tabularx}{0.49\textwidth}{|>{\RaggedRight\arraybackslash}X|}
      \hline
      \\
      \normalsize
        \tool can identify CWEs with a maximum of 68\% accuracy for perfect CWE matches, 68.6\% accuracy for prediction coverage, and 69.2\% accuracy for ground truth coverage, outperforming the baseline experiments.\\
      \\
      \hline
    \end{tabularx}
\end{table}

\begin{figure}[t]
\centering
\includegraphics[width=0.5\textwidth]{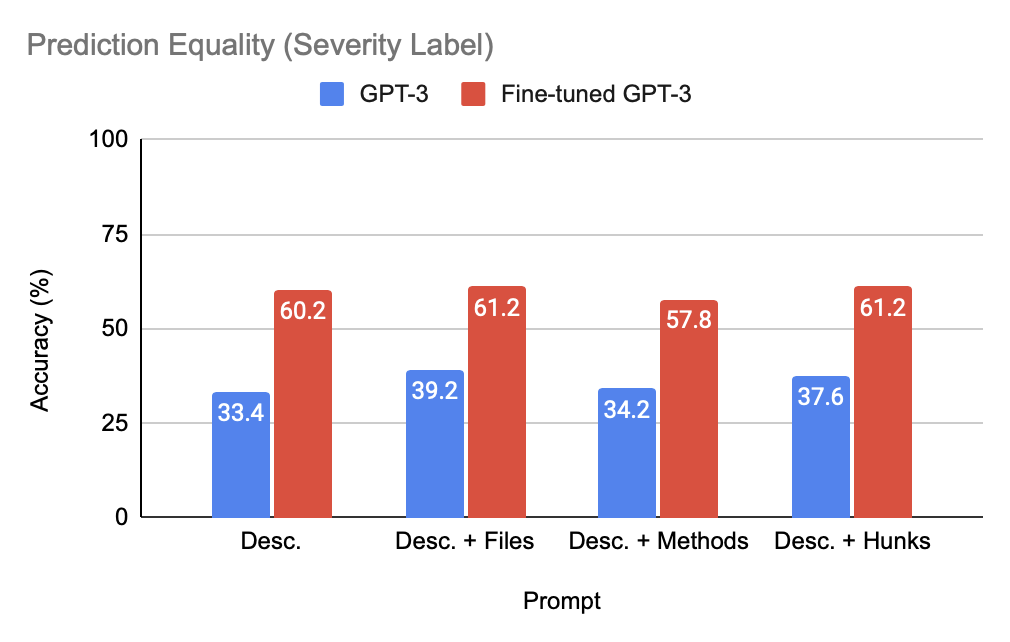}
\caption{\label{fig:severity_label}Prediction Equality (Severity Label).}
\end{figure}

\begin{figure}[t]
\centering
\includegraphics[width=0.5\textwidth]{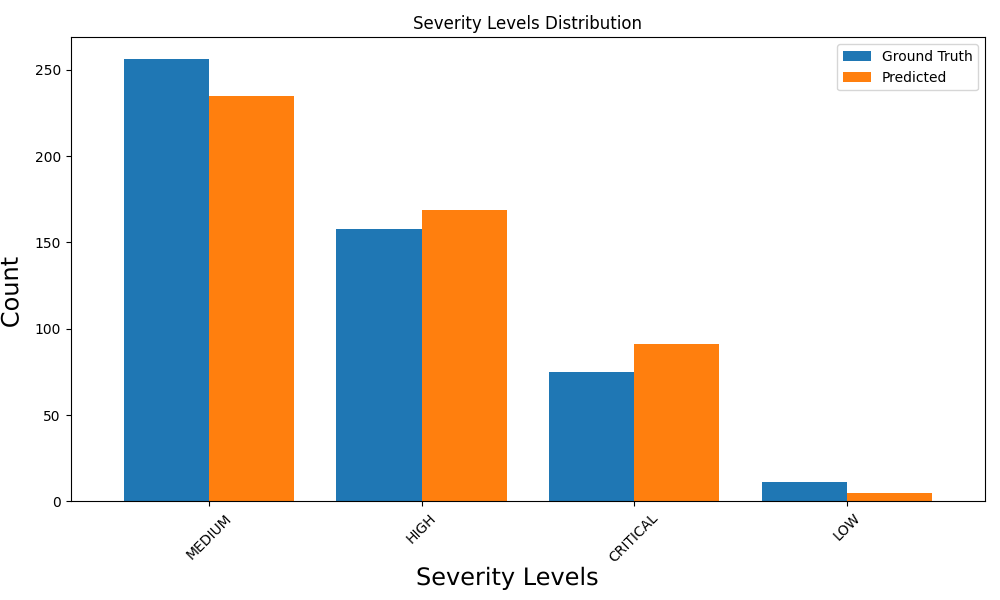}
\caption{\label{fig:severity_analysis_severity_distribution_DH_fine}Distribution of Severity Label Identification (Fine-tuned GPT-3 Providing Description + Hunks).}
\end{figure}

\begin{figure}[t]
\centering
\includegraphics[width=0.5\textwidth]{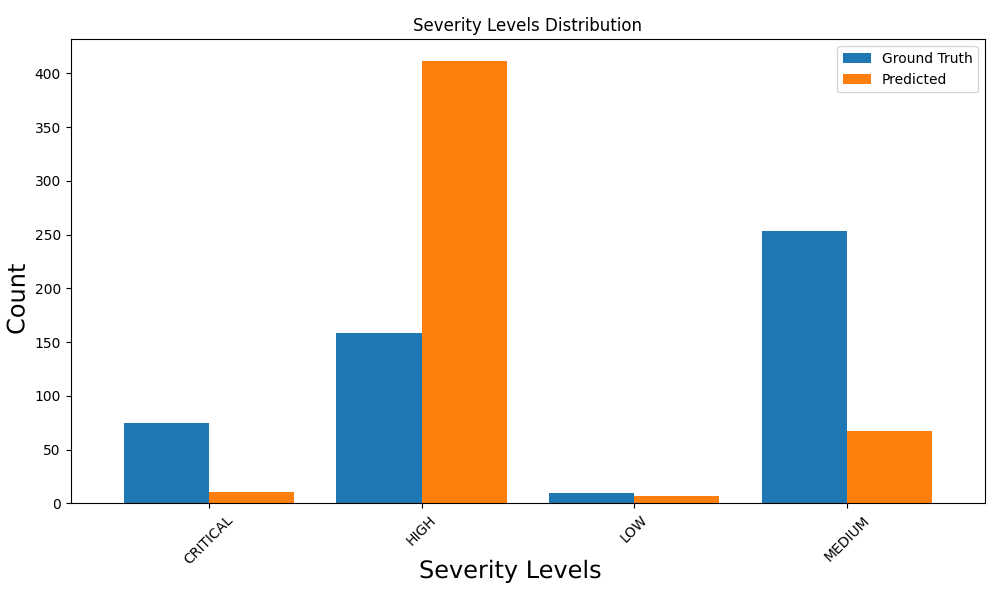}
\caption{\label{fig:severity_analysis_severity_distribution_DH_3}Distribution of Severity Label Identification (Baseline Providing Description + Hunks).}
\end{figure}

\textbf{RQ2: Can \tool accurately identify severities based on their associated CVSS version?} We categorize the severity evaluation into severity score and severity label as follows:

\textbf{Severity Label:} Figure \ref{fig:severity_label} shows the accuracy of the tool in terms of identifying the severity labels that are equal to the ground truth according to their corresponding CVSS. According to this figure, the tool performs better than the baseline experiment, with accuracy ranging from 57.8\% to 61.2\%. Despite the CWE identifications, providing the buggy source files along with the bug description improves the accuracy in terms of identifying severity labels. The first variant (providing a description) had a maximum difference of 26.8\% between adopting the fine-tuned model and the baseline experiment. Our experiments also show that adopting the fine-tuned model by \tool results in a more equitable distribution in the identification of severity labels. An example of this claim is shown in Figure \ref{fig:severity_analysis_severity_distribution_DH_fine}, where the identified labels (orange) match the ground truth labels (blue). Comparing this Figure with Figure \ref{fig:severity_analysis_severity_distribution_DH_3} highlights the effectiveness of our approach when adopting the fine-tuned model to better capture the nuances of severity levels, reducing incorrect label identifications. This consistency and alignment between identified and ground truth labels demonstrate the robustness and reliability of the tool when adopting the fine-tuned model \footnote{This pattern is also evident across other cases within as diagrams in the released artifacts.}.

\begin{figure}[t]
\centering
\includegraphics[width=0.5\textwidth]{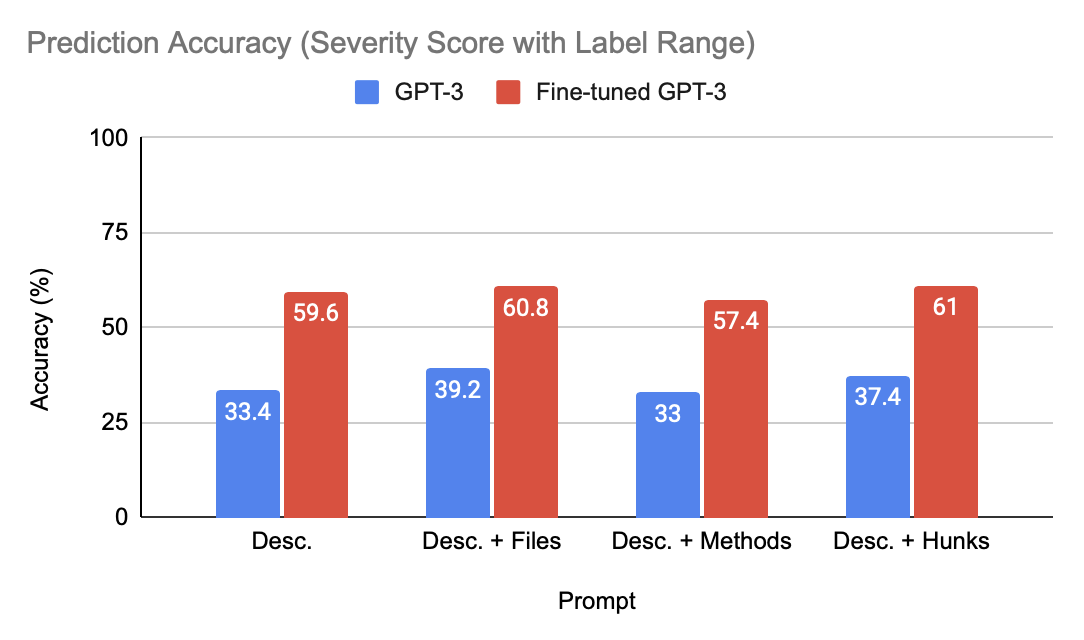}
\caption{\label{fig:severity_score_with_label_range}Prediction Accuracy (Severity Score - Label Range).}
\end{figure}

\begin{figure}[t]
\centering
\includegraphics[width=0.5\textwidth]{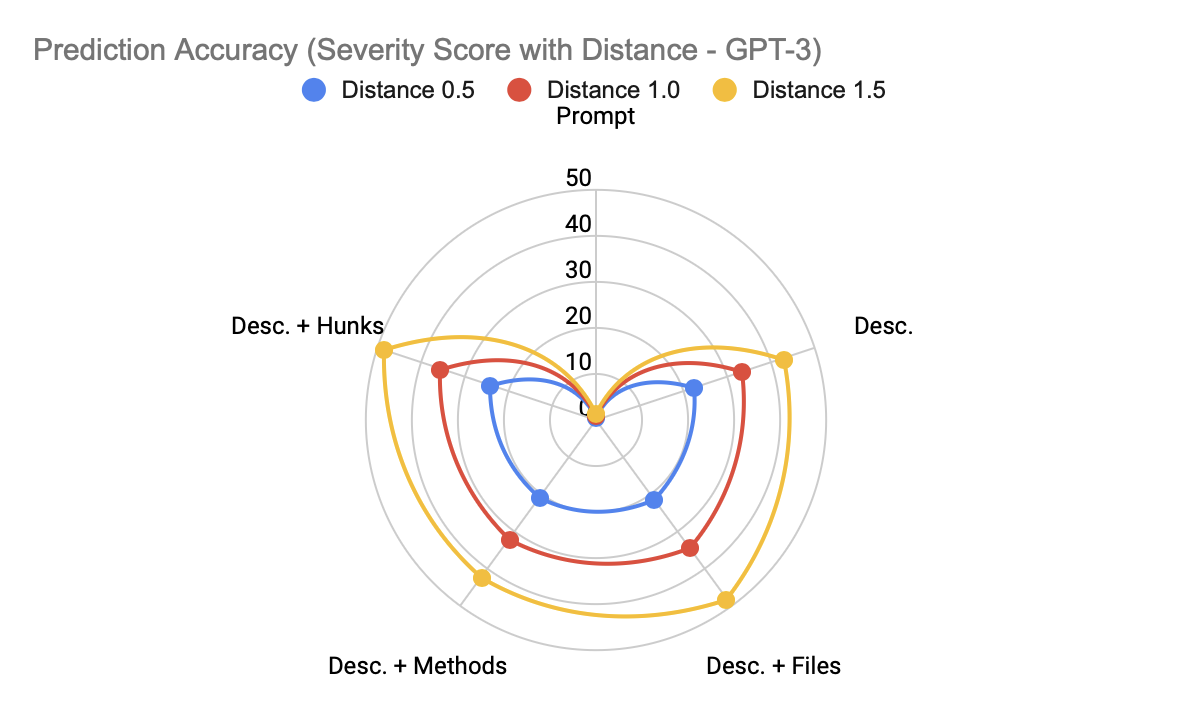}
\caption{\label{fig:severity_score_distances}Prediction Accuracy (Severity Score - Distance).}
\end{figure}

\begin{figure}[t]
\centering
\includegraphics[width=0.5\textwidth]{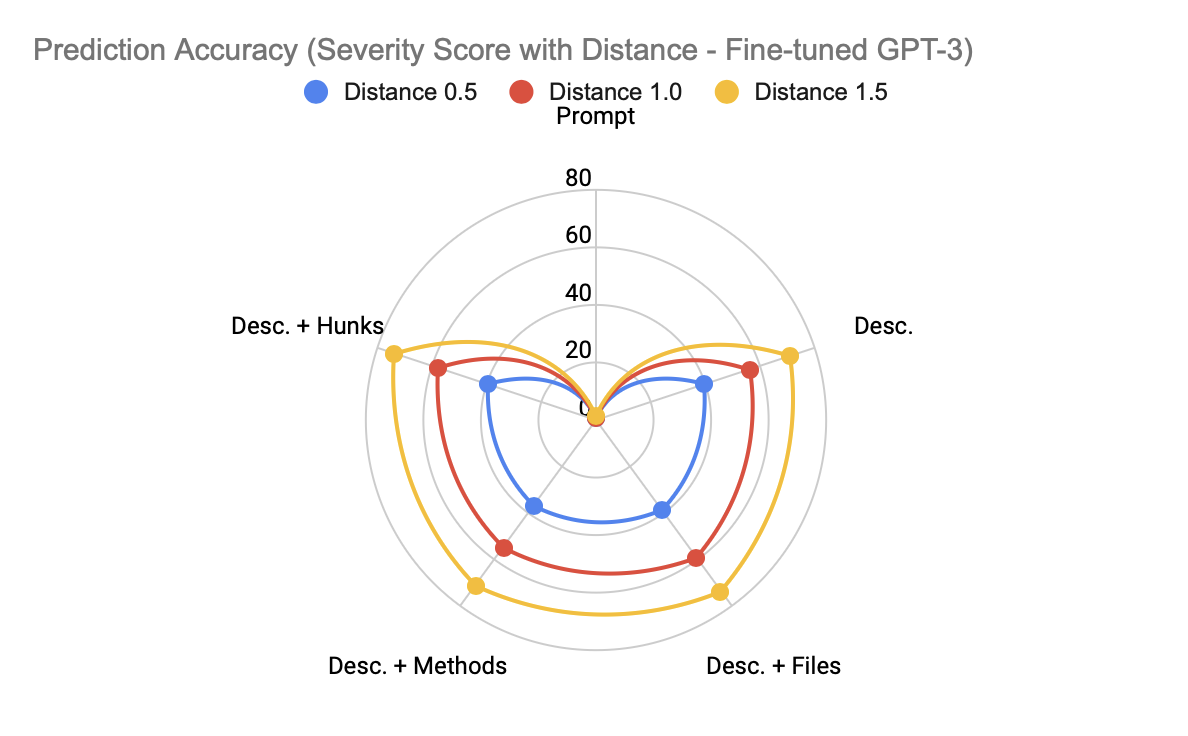}
\caption{\label{fig:severity_score_distance_finetuned_gpt3}Prediction Accuracy (Severity Score with Distance).}
\end{figure}

\textbf{Severity Score:} 
Figure \ref{fig:severity_score_with_label_range} shows the accuracy of \tool in terms of generating severity scores whose enclosing labels are equal to the ground truth severity labels. Comparing this figure with Figure \ref{fig:severity_score_with_label_range}, the results show that the accuracies are nearly identical. This indicates that the identified severity labels are often consistent with the label range of the associated identified severity scores. Another criterion to measure the accuracy of the severity score is the distance, with results depicted in Figures \ref{fig:severity_score_distances} and \ref{fig:severity_score_distance_finetuned_gpt3}. Overall, the results demonstrate that our approach, when adopting the fine-tuned model (Figure \ref{fig:severity_score_distance_finetuned_gpt3}), outperforms the baseline experiment (Figure \ref{fig:severity_score_distances}). According to Figure \ref{fig:severity_score_distances}, \tool performs maximally (considering a distance of 1.5) when using the ``description + hunks'' prompts. The accuracy is 48.6\% in the baseline and 73.6\% with the fine-tuned model. As we increase the distance from 0.5 to 1.5, the accuracy of the tool increases since it can cover a broader range of severity scores.

\begin{table}[t]
    \label{box:rq2_summary}
    \begin{tabularx}{0.49\textwidth}{|>{\RaggedRight\arraybackslash}X|}
      \hline
      \\
      \normalsize
        \tool can identify severities with maximum accuracies of 61.2\% for severity label identification, 61\% for severity score (label range), and 73.6\% for severity score (distance 1.5), outperforming the baseline experiments.\\
      \\
      \hline
    \end{tabularx}
\end{table}

\textbf{RQ3: Can \tool accurately identify CWEs and severities collectively?} Overall, our experiments demonstrate a roughly threefold increase in \tool's accuracy when adopting our fine-tuned model compared to the baseline. 

\begin{figure}[t]
\centering
\includegraphics[width=0.5\textwidth]{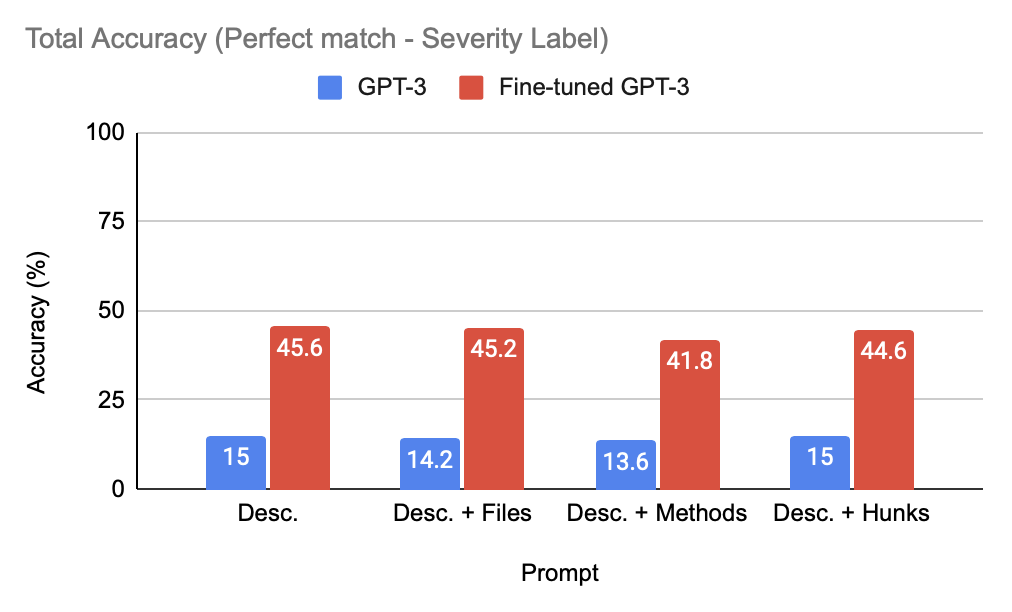}
\caption{\label{fig:overall_equality_label}Total Accuracy (Perfect Match - Severity Label).}
\end{figure}

We demonstrate the effectiveness of our approach using different criteria as follows:

\textbf{Perfect Match + Severity Label:} This criterion evaluates the accuracy in identifying correct severity labels and perfect matches for CWEs. As shown in Figure \ref{fig:overall_equality_label}, \tool achieves its highest accuracy of 45.6\% with our fine-tuned model when including the bug description in the prompt, while the lowest accuracy of 41.8\% is observed when including the bug description and buggy methods. For the ``description + files'' variant, compared to the baseline experiments, there is a 31\% increase in \tool's accuracy, highlighting the effectiveness of our contribution.

\begin{figure}[t]
\centering
\includegraphics[width=0.5\textwidth]{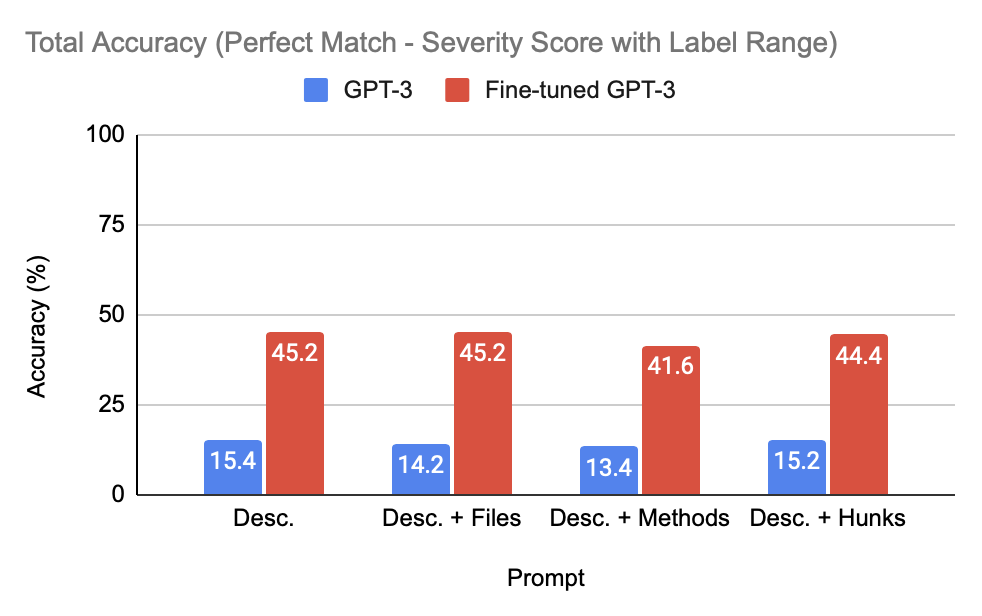}
\caption{\label{fig:overall_equality_label_range}Total Accuracy (Perfect Match - Severity Score - Label Range).}
\end{figure}

\textbf{Perfect Match + Severity Score with Label Range:} This criterion evaluates the accuracy of the tool in identifying perfect match CWEs as well as severity scores whose label range matches the ground truth severity score's label range. As depicted in Figure \ref{fig:overall_equality_label_range}, \tool performs best when provided with the bug description optionally with the buggy files, achieving an accuracy of 45.2\%. Conversely, the lowest accuracy across the four prompts is observed when using the ``description + methods'' prompt (41.6\%). Regarding the baseline experiments, the accuracy is roughly the same as the previous criterion.

\begin{figure}[t]
\centering
\includegraphics[width=0.5\textwidth]{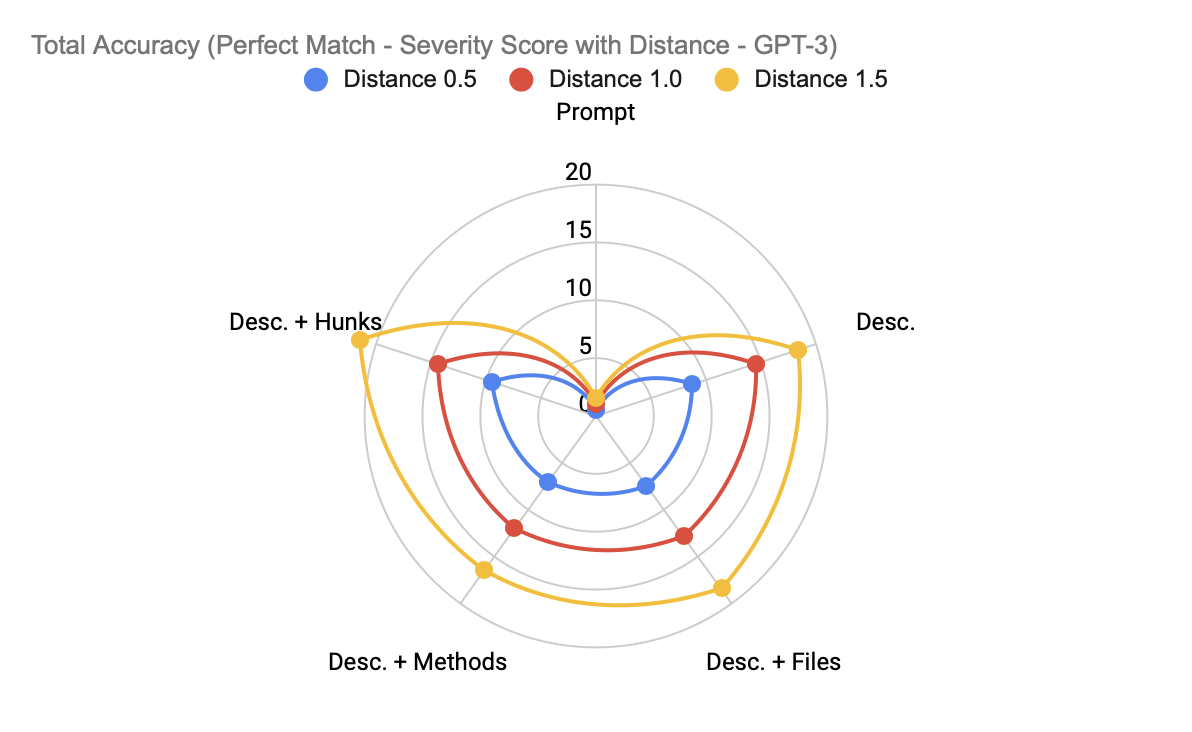}
\caption{\label{fig:overall_perfect_match_distance_gpt3-gpt3}Total Accuracy (Perfect Match - Severity Score - Distance).}
\end{figure}

\begin{figure}[t]
\centering
\includegraphics[width=0.5\textwidth]{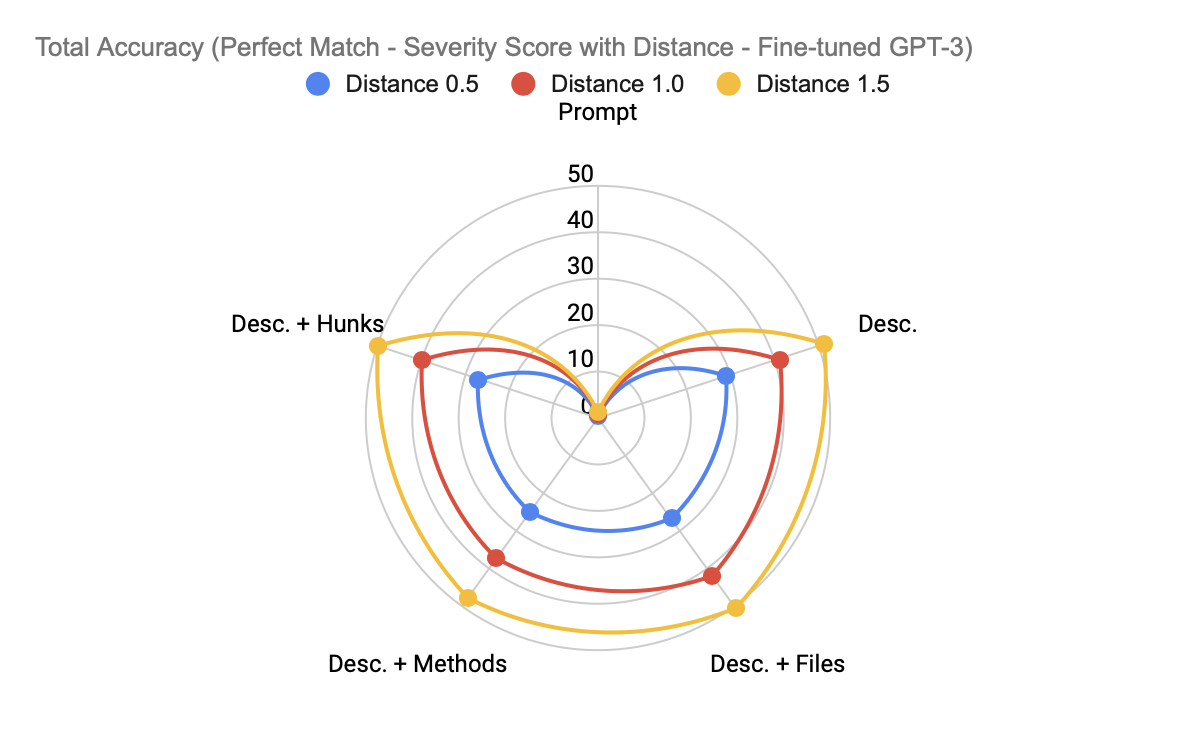}
\caption{\label{fig:overall_perfect_match_distance_finetuned_gpt3-finetuned-gpt3}Total Accuracy (Perfect Match - Severity Score - Distance).}
\end{figure}

\textbf{Perfect Match + Severity Score with Distance:} This criterion evaluates the accuracy of \tool in identifying correct perfect match CWEs as well as severity scores that align with the ground truth severity score within specific distances (0.5, 1.0, and 1.5). As shown in Figure \ref{fig:overall_perfect_match_distance_gpt3-gpt3}, \tool's accuracy ranges from 47.8\% to 51.2\% when using the ``description + methods'' and bug description for inference, respectively. Compared to the baseline results depicted in Figure \ref{fig:overall_perfect_match_distance_finetuned_gpt3-finetuned-gpt3}, these findings highlight the effectiveness of our approach when adopting our fine-tuned model over the baseline experiment.

\begin{figure}[t]
\centering
\includegraphics[width=0.5\textwidth]{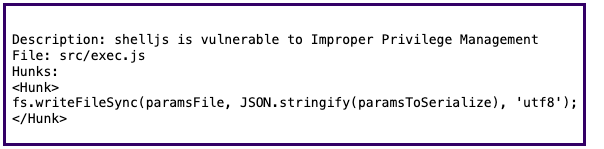}
\caption{\label{fig:nuance_difference}Input Example Where Fine-tuning Improves.}
\end{figure}

To exemplify, consider the input prompt shown in Figure \ref{fig:nuance_difference}, extracted from the ``description + hunks'' variant. We observed that in baseline, the tool mistakenly identified CWE-266 (Incorrect Privilege Assignment) as the correct match CWE. However, when adopting our fine-tuned model, it correctly identified CWE-269 (Improper Privilege Management), which matches the ground truth CWE. This demonstrates the effectiveness of fine-tuning in better grasping the nuances in the bug description and the provided hunk, resulting in accurate CWE identification.

\begin{table}[h]
    \label{box:rq3_summary}
    \begin{tabularx}{0.49\textwidth}{|>{\RaggedRight\arraybackslash}X|}
      \hline
      \\
      \normalsize
     \tool achieves its performance in CWE and severity identification with a maximum accuracy of 51.2\%, by correctly identifying perfect CWE matches and severity scores within a 1.5 distance, thereby outperforming the baseline experiment.\\
      \\
      \hline
    \end{tabularx}
\end{table}

{\textbf{Key Insights:}}
The combined results of all experiments demonstrate the effectiveness of our approach in identifying CWEs and severities and their combination together. The findings highlight the importance of bug descriptions in achieving higher accuracy. In contrast, relying solely on buggy code at either granularity level results in the lowest accuracy, confirming that additional contextual information is beneficial rather than solely relying on providing buggy codes. For experiments involving both code and descriptions, the results indicate that the combination of hunk-level granularity and bug descriptions achieves the highest accuracy for both CWE and severity identification. This outcome can be attributed to the tool receiving only the necessary context to generate CWEs (as either perfect matches or top candidates) and severities (as labels or scores). Conversely, providing the tool with entire buggy files leads to the lowest accuracy, likely due to the tool encountering excessive information and noise. Typically, only a few lines in the file are edited to fix the bug, while the remaining content misleads the tool. In summary, the results reveal that incorporating bug descriptions—either alone or in combination with buggy hunks—yields the most effective configurations for maximizing accuracy. For CWE identification, the tool tends to overpredict perfect match CWEs. For severity identification, considering severity labels or severity scores with a distance metric helps improve accuracy. These insights underscore the importance of providing focused and relevant context to the tool (as we did in four variants), minimizing noise, and enhancing its predictive performance.

\subsection{\textbf{Manual Analysis:}} By taking a detailed look at the results where the identified and ground truth CWEs are not identical, we observed instances where the tool identified a more precise CWE than the ground truth, demonstrating \tool's ability to pinpoint finer distinctions within the CWE hierarchy. For example, in one case, the ground truth perfect match CWE was CWE-94 (Improper Control of Generation of Code - Code Injection), a broad vulnerability related to insecure code generation. However, the tool identified CWE-95 (Improper Neutralization of Directives in Dynamically Evaluated Code - Eval Injection), which is a child of CWE-94 in the official CWE hierarchy. This specific classification reflects the tool's capacity to correctly identify vulnerabilities at a more granular level, highlighting its effectiveness in distinguishing between related but distinct issues. Conversely, we also found major cases where the tool identified a parent CWE rather than a child CWE. For example, in one instance, the ground truth was labeled as CWE-78 (Improper Neutralization of Special Elements used in an OS Command - OS Command Injection), a more specific vulnerability related to command injection. However, the tool identified CWE-77 (Improper Neutralization of Special Elements used in a Command - Command Injection), a parent of CWE-78 in the official CWE hierarchy.

As another manual analysis, we selected the ``description + hunks'' variant and randomly sampled 50 records along with their evaluation results generated by the tool. This variant was chosen because it includes the buggy code at the finest granularity level, and enables us to assess the tool's performance across different programming languages. The results are summarized in Table \ref{tab:manual_analysis}, wherein \texttt{Alpha} refers to the number of records (out of 50) for which the perfect-match CWEs were correctly identified by the tool, and \texttt{Beta} represents the number of records (out of 50) for which the severity labels were correctly identified by the tool. The tool correctly identified the CWE with a perfect match in 35 cases, aligning with the ground truth CWEs. For severity labels, 33 cases had matches with the expected ground truth. Examining performance across programming languages, PHP showed the highest representation, with 12 records, of which 10 had correct CWE (\texttt{Alpha} = 10) and severity label (\texttt{Beta} = 10) identifications. Next, the C programming language stands with 7 records, of which 5 had correct CWE identifications (\texttt{Alpha} = 5) and 3 had correct severity label identifications (\texttt{Beta} = 3). These two languages accounted for the most frequently occurring records in the sampled data. They show roughly 83.3\% and 71.4\% accuracy for CWE identification, respectively, and 82.3\% and 42.8\% accuracy, respectively, for severity label identification. The lowest-performing language was C++, with only one record, where neither the CWE nor the severity label was correctly identified.

\begin{table}
\caption{Correct Identifications by Language}
\label{tab:manual_analysis}
\centering
\normalsize
\begin{tabular}{|c|c|c|c|}
\hline
\textbf{Programming Language} & \textbf{\#Records} & \textbf{Alpha} & \textbf{Beta} \\
\hline
PHP       & 12 & 10 & 10 \\
\hline
C         & 7  & 5  & 3 \\
\hline
JavaScript& 5  & 5  & 4 \\
\hline
Java      & 5  & 3  & 4 \\
\hline
TypeScript & 4  & 2  & 3 \\
\hline
Python    & 4  & 3  & 2 \\
\hline
Go        & 2  & 2  & 2 \\
\hline
C++       & 1  & 0  & 0 \\
\hline
\end{tabular}
\end{table}

\subsection{\textbf{Failed Case Analysis:}}
\begin{figure}[h!]
\centering
\includegraphics[width=0.5\textwidth]{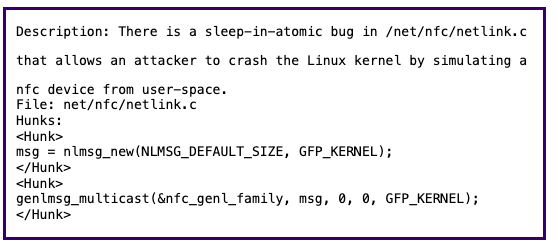}
\caption{\label{fig:failed_sample}\tool Sample Input For Incorrect Identification.}
\end{figure}

Our analysis also reveals that one of the key challenges for \tool in accurately predicting CWEs is the lack of sufficient and rich information provided about the bug, either as description, code, or both combined. Despite the strength of the fine-tuned model in learning patterns from the training dataset, the CWE predictions sometimes fail due to flawed or ambiguous input information. Conversely, providing the tool with the entire buggy files introduces an overload of information, which can also lead to incorrect identifications of CWEs and severity levels. An example of this limitation is shown in Figure \ref{fig:failed_sample}, where the tool was given an incomplete bug description and a buggy hunk as input. Consequently, the tool incorrectly inferred CWE-20 (Improper Input Validation) instead of the ground truth, CWE-248 (Uncaught Exception). This highlights the tool's dependence on critical contextual details, such as the code's intent and its interaction with other components.

Our analysis also identifies other causes of incorrect identifications. One such reason is the lack of relevant information about certain bugs in the fine-tuning dataset. Since the fine-tuning dataset is distinct from the evaluation dataset, it may not adequately represent all the bugs that the tool encounters during evaluation. Moreover, in some cases, the irrelevant or misleading information within the inputs can misguide the tool, further lowering its accuracy. Additionally, a recurring reason for failure is the misclassification of sibling, child, or parent CWEs instead of the expected CWE. While these identifications are related, they are still considered failures because they do not match the ground truth perfect match. Exploring the relationships between expected and identified CWEs presents an avenue for future work to refine the evaluation criteria. Finally, the inherent limitations of the language model including the hallucinations and the model's architectural constraints (like input token size) also contribute to failure. Updating to newer versions of the model, provided they adhere to a static cut-off date such that the fine-tuning and evaluation datasets are all unseen, could potentially enhance accuracy and mitigate such issues.

\subsection{\textbf{Non-NVD Dataset Analysis:}}
Regarding evaluating the \tool performance for other datasets, we fed the tool with five non-NVD evaluation records each having descriptions and buggy codes at file, method, and hunk granularity levels. These records were newly discovered CVEs (published in 2025) that were reported by VulDB\footnote{https://vuldb.com} and CVEDetails\footnote{http://cvedetails.com} open-source repositories and were not publicly available in the NVD at the time of this evaluation. By executing the tool across all variants, three records were discarded due to the prompt token exceedance constraint imposed by the language model. For the remaining, across all four experimental variants, the tool when adopting the fine-tuned model produced correct perfect match CWEs, whereas the standard model (as baseline) did not. Therefore, these results, although being conducted in small size, highlight the potential effectiveness of our proposed approach in terms of CWE identification. Regarding severity analysis, improvements were observed in the results as in some variants (prompts), the identified labels matched the ground truth, and the severity scores were closely aligned. Again, given that the non-NVD dataset size was smaller compared to the collected NVD dataset, we should look at these results with caution. In conclusion, the results show our tool's potential effectiveness in working with any datasets as long as they provide the tool with required inputs and follow the input formats described within this paper.

\section{Threats to validity}
\label{sec:threats_to_validity}

While our study addresses key challenges in CWE and severity identification, several threats remain. First, we relied on NVD descriptions, which are often template-generated, limiting the tool to fully leverage the capabilities of LLMs. Regarding the dataset, although our dataset covers various programming languages and software projects, we excluded some reports due to unsupported programming language, token limitations, and the unavailability of specific buggy code snippets. To mitigate, we utilized the maximum allowed number of tokens during inference.
\section{Related Work}
\label{sec:related_works}

Our research distinguishes itself by specifically focusing on automating bug report categorization through the use of Large Language Models. This section reviews some relevant studies that address bug localization, triage, and priority assessment. 

\paragraph{Automated Association of Security Knowledge Bases} 
Giannakopoulos and Maliatsos \cite{giannakopoulos2023usage} trained a text classification model on a privately labeled dataset of CVE descriptions and attack patterns to predict and extrapolate threats. Unlike our approach which identifies CWEs and severity using LLMs, their approach employs a text classification model trained on a dataset mapping vulnerability descriptions to threat descriptions. In another study, Bonomi et al. \cite{bonomi2025surfacenlpbasedmethodologyautomatically} proposed a hybrid NLP methodology that associates CVEs with Common Attack Pattern Enumeration and Classification attack patterns, aiming to correlate vulnerabilities with attack strategies for threat analysis and attack pattern ranking. Their method integrated semantic similarity (embedding vectors) and keyword search to enhance the accuracy of these associations. While their work centers on linking vulnerabilities to attack patterns, our research emphasizes automating CWE and severity extraction. This highlights the broad field of security vulnerability analysis to which both approaches contribute, offering complementary solutions for automating critical aspects of vulnerability classification and threat prioritization.

\paragraph{Bug Triage} Zexuan et al. \cite{6227209} proposed a socio-technical model to rank developers based on contributions, enhancing bug triage and severity identification tasks. Their approach achieved a 13\% improvement in bug triage accuracy by prioritizing developers. They prioritized developers, which is assistive in bug triaging. While their approach performs well in Mozilla and Eclipse projects, our approach considers a broader range of real-world projects. 

\paragraph{Bug Localization} Richter and Wehrheim \cite{10298391} studied neural bug detectors by comparing models trained on real versus synthetic bugs, demonstrating the impact of real-world bug fixes for localization and repair tasks. In another study, Niu et al. \cite{10172549} introduced RAT, a traceability model that integrates refactoring data to improve bug localization, particularly in refactoring-intensive projects. Unlike our approach that uses LLMs, this research enhances localization through code evolution tracking requiring historical code changes.

\paragraph{Bug Classification} Mills et al. \cite{8530045} evaluated text retrieval techniques for bug localization, proposing a genetic algorithm to improve query formulation. While this method relies on optimizing search queries, our approach automates bug-type tagging by using a pre-trained LLM, which moves beyond retrieval to interpret context for more detailed bug classification. In another study, Izadi et al. \cite{10.1007/s10664-021-10085-3} employed the fine-tuned RoBERTa model in a two-stage approach to classifying issue objectives and priorities, achieving 82\% accuracy in objective prediction and 75\% in priority. Similarly, we use GPT-3 for bug tagging but go further to categorize bugs by type and severity, adding classification specificity.

\section{Conclusion}
\label{sec:conclusion}

We introduced \tool, a novel approach that leverages Large Language Models to automate the identification of Common Weakness Enumerations and the assessment of severities for security vulnerabilities. By utilizing prompt engineering techniques, \tool achieved a CWE identification accuracy of 68\%, a severity identification accuracy of 73.6\%, and 51.2\% combined. These results highlight the potential of our approach to streamline software vulnerability management and enhance the efficiency of bug triaging workflows. In the future, we plan to evaluate our approach using other open-source and smaller LLMs and further assess its generalizability.
\section*{Data Availability}
\label{sec:data_availability}

All materials and artifacts associated with this study are publicly available at \url{https://zenodo.org/records/14776104}.


\bibliographystyle{IEEEtran}
\bibliography{list}

\newpage

\appendices
\section{}
\label{sec:appendix_a}

\end{document}